\documentclass[lettersize,journal]{IEEEtran}

\usepackage{amsmath,amsfonts}
\usepackage{bbold}
\usepackage{cite}
\usepackage{array}
\usepackage{textcomp}
\usepackage{stfloats}
\usepackage{url}
\usepackage{verbatim}
\usepackage{graphicx}  
\usepackage{subcaption}
\usepackage{caption}
\usepackage{textcomp}
\usepackage{xcolor}
\usepackage{multirow}
\usepackage{eurosym}
\usepackage{dirtytalk}
\usepackage{hyperref}
\usepackage[linesnumbered,ruled]{algorithm2e}
\allowdisplaybreaks

\usepackage{scalerel}

\newcommand{\Rb}{\ensuremath{\mathbb{R}}}

\newcommand{\GG}{\ensuremath{\mathcal{G}}}
\newcommand{\VV}{\ensuremath{\mathcal{V}}}
\newcommand{\EE}{\ensuremath{\mathcal{E}}}
\newcommand{\pinj}{\ensuremath{P^{\mathtt{inj}}}}
\newcommand{\qinj}{\ensuremath{Q^{\mathtt{inj}}}}

\usepackage{amssymb}
\usepackage{nomencl}
\makenomenclature

\usepackage{etoolbox}
\renewcommand\nomgroup[1]{%
  \item[\bfseries
  \ifstrequal{#1}{P}{\textit{C. Variables}}{%
  \ifstrequal{#1}{N}{\textit{A. Sets}}{%
  \ifstrequal{#1}{O}{\textit{B. Parameters}}{}}}%
]}

\newcommand{\ignore}[1]{}

\begin{document}

\title{Prosumer-Centric Flexible Dynamic Operating Envelopes for Low-Voltage Distribution Networks} 

\author{Abhishek Mishra,~\IEEEmembership{Graduate Student Member,~IEEE,} Ashish R. Hota,~\IEEEmembership{Senior Member,~IEEE,} Gayan Lankeshwara,~\IEEEmembership{Member,~IEEE,} Rahul Sharma,~\IEEEmembership{Senior Member,~IEEE,} Wayes Tushar,~\IEEEmembership{Senior Member,~IEEE,} Prabodh Bajpai,~\IEEEmembership{Senior Member,~IEEE} 

\thanks{A. Mishra and A. R. Hota are with the Department of Electrical Engineering, Indian Institute of Technology Kharagpur, West Bengal, India. Email: abhishekmishra.ee21{@}kgpian.iitkgp.ac.in, ahota{@}ee.iitkgp.ac.in. (Corresponding author: Abhishek Mishra.)

G. Lankeshwara, R. Sharma and W. Tushar are with the School of Electrical Engineering and Computer Science, The University of Queensland, Brisbane, QLD 4072, Australia. Email: g.lankeshwara{@}uq.edu.au, rahul.sharma{@}uq.edu.au, w.tushar{@}uq.edu.au.

P. Bajpai is with the Department of Sustainable Energy Engineering, Indian Institute of Technology Kanpur, Uttar Pradesh, India. Email: pbajpai{@}iitk.ac.in}}%

\maketitle
\begin{abstract}
As the share of distributed energy resources (DERs) increases in the distribution network, maintaining compliance with network constraints has become a challenge. In this context, dynamic operating envelopes (DOEs) have emerged as a promising solution, where the distribution network operator (DNO) computes and imposes a dynamically varying import and export limit on power exchange between prosumers and the distribution network. Existing DOE approaches often require the prosumers to report their desired power exchange to the DNO, which then computes DOE limits that typically do not exceed the reported values. However, due to uncertain renewable generation and load demand, such DOE limits can potentially result in unnecessary curtailment of generation and load during real-time operation. This work addresses this limitation by computing flexible DOEs using a flexible optimization framework that trades off optimality with flexibility. The proposed approach computes both upper and lower limits on the active and reactive power exchange between each prosumer and the grid, and tries to contain the reported values between the upper and lower limits. As long as the power exchange resides within the DOE limits, network constraints are satisfied. The proposed flexible DOE is validated on a modified Australian low-voltage distribution network. Compared to non-flexible DOE, the proposed framework demonstrates superior performance in reducing curtailment and total operational costs while consistently maintaining voltage magnitudes within desired limits.
\end{abstract}


\nomenclature[P]{\(P^{g}_{b,t}/Q^{g}_{b,t}\)}{Active/reactive power purchased from the grid.}
\nomenclature[P]{\(P^{g}_{s,t}/Q^{g}_{s,t}\)}{Active/reactive power sold to the grid.}
\nomenclature[P]{\(\pinj_{t}/\qinj_{t}\)}{Active/reactive power injection into ADN by the prosumer.}
\nomenclature[P]{\(e_{i,j,t}\)}{Energy stored in the $j^{\text{th}}$ segment of the linearized BESS degradation model of prosumer $i$ at time $t$.}
\nomenclature[P]{\(P^{\mathtt{ch}}_{t}/P^{\mathtt{dis}}_{t}\)}{Charging/discharging power of the BESS.}
\nomenclature[P]{\(p^{\mathtt{ch}}_{i,j,t}/p^{\mathtt{dis}}_{i,j,t}\)}{Charging/discharging power of the $j^{\text{th}}$ segment of the linearized BESS degradation model of prosumer $i$ at time $t$.}
\nomenclature[P]{\(P^{\mathtt{inv}}_{t}/Q^{\mathtt{inv}}_{t}\)}{Active/reactive power output of the inverter.}
\nomenclature[P]{\(\delta_{i,t}\)}{Binary variable that reflects whether prosumer $i$ is a buyer or seller at time $t$.}
\nomenclature[P]{\(P^{\mathtt{inj},\mathtt{DNO}}_{t}/Q ^{\mathtt{inj},\mathtt{DNO}}_{t}\)}{Nodal active/reactive power injection specified by the DNO.}
\nomenclature[P]{\(P^j_{t}/Q^j_{t}\)}{Active/reactive power flow in ADN branches.}
\nomenclature[P]{\(I^{\mathtt{sq},j}_{t}\)}{Magnitude of squared current in branch $j$ at time $t$.}
\nomenclature[P]{\(V^{\mathtt{sq}}_{j,t}\)}{Magnitude of squared voltage at node $j$ at time $t$.}
\nomenclature[P]{\(\mathbf{x}_{t}\)}{Nodal power injection vector at time $t$.}
\nomenclature[P]{\(\beta_t\)}{Flexibility variables at time $t$.}
\nomenclature[P]{\(\mathbf{z}\)}{Uncertain variables.}
\nomenclature[P]{\(P^{\mathtt{PV}}_{\mathtt{curt},t}/P^{\mathtt{load}}_{\mathtt{curt},t}\)}{Curtailment in PV/load.}
\nomenclature[P]{\(\mathbf{x}_{t}^\star/\beta_t^\star\)}{Optimal power exchange/flexibility computed by DNO at time $t$.}
\nomenclature[P]{\(\theta_t\)}{Cycle depth.}
\nomenclature[P]{\(P^{\mathtt{DIS}}_t\)}{Discharge power.}
\nomenclature[P]{\(\gamma_t\)}{Cycle depth stress function.}
\nomenclature[P]{\(\alpha_{i,t}\)}{Binary variable to prevent simultaneous charging/discharging of the BESS of prosumer $i$ at time $t$.}
\nomenclature[P]{\(\)}{}
\nomenclature[P]{\(\)}{}
\nomenclature[P]{\(\)}{}
\nomenclature[P]{\(\)}{}
\nomenclature[P]{\(\)}{}
\nomenclature[P]{\(\)}{}
\nomenclature[P]{\(\)}{}
\nomenclature[P]{\(\)}{}

\nomenclature[O]{\(\eta_i^{\mathtt{ch}}/\eta_i^{\mathtt{dis}}\)}{BESS charging/discharging efficiency.}
\nomenclature[O]{\(\tau \)}{Sample time.}
\nomenclature[O]{\(N_p\)}{Prediction horizon.}
\nomenclature[O]{\(n\)}{Number of prosumers.}
\nomenclature[O]{\(\pi ^{\mathtt{ToU}}_t/\pi ^{\mathtt{FiT}}_t\)}{Time-of-use tariff/Feed-in-tariff.}
\nomenclature[O]{\(c^{\mathtt{PV}}_t/c^{\mathtt{load}}_t\)}{PV/load curtailment cost.}
\nomenclature[O]{\(c^{\mathtt{bat}}\)}{Penalty for deviation from optimal SOC.}
\nomenclature[O]{\(P^{\mathtt{PV},\mathtt{frc}}_{i,t}/P^{\mathtt{PV},\mathtt{unc}}_{i,t}\)}{Forecasted/realized PV power output.}
\nomenclature[O]{\(E_{\min,i}/E_{\max,i}\)}{Min/max energy capacity limit of BESS.}
\nomenclature[O]{\(\bar{e}_{i,j}\)}{Maximum energy associated with the $j^{\text{th}}$ segment of the piecewise linear degradation model.}
\nomenclature[O]{\(P^{\mathtt{bat}}_{\max,i}\)}{Max BESS charging/discharging power.}
\nomenclature[O]{\(S^{\mathtt{inv}}_{i}\)}{Inverter apparent power rating.}
\nomenclature[O]{\(P^{\mathtt{load},\mathtt{frc}}_{i,t}/Q^{\mathtt{load},\mathtt{frc}}_{i,t}\)}{Active/reactive forecasted load demand.}
\nomenclature[O]{\(P^{\mathtt{load},\mathtt{unc}}_{i,t}/Q^{\mathtt{load},\mathtt{unc}}_{i,t}\)}{Active/reactive realized load demand.}
\nomenclature[O]{\(P^{\mathtt{inj}}_{\min,i}/P^{\mathtt{inj}}_{\max,i}\)}{Min/max active power injection limit at the prosumer nodes.}
\nomenclature[O]{\(Q^{\mathtt{inj}}_{\min,i}, Q^{\mathtt{inj}}_{\max,i}\)}{Min/max reactive power injection limit at the prosumer nodes.}
\nomenclature[O]{\(a_j^p, b_j^q, c_j^s\)}{Coefficients of the linearized inverter constraints}
\nomenclature[O]{\(r_j,x_j\)}{Resistance/reactance of ADN line $j$.}
\nomenclature[O]{\(\bar{I}^{\mathtt{sq},j}_{t},\bar{P}^j_{t}/\bar{Q}^j_{t},\bar{V}^{\mathtt{sq},j}_{t}\)}{Initial estimate of square of line current, active/reactive line flow and square of nodal voltage, respectively.}
\nomenclature[O]{\(A^1_t, A^2_t, A^3_t,b^1_t, b^2_t, b^3_t\)}{Coefficients of power flow equations.}
\nomenclature[O]{\(\mathbf{y}^{\mathtt{ref}}_{i,t}\)}{Desired active power exchange of the prosumer.}
\nomenclature[O]{\(P^{\mathtt{DOE},+}_t/Q^{\mathtt{DOE},+}_t\)}{Active/reactive power upper DOE limit.}
\nomenclature[O]{\(P^{\mathtt{DOE},-}_t/Q^{\mathtt{DOE},-}_t\)}{Active/reactive power lower DOE limit.}
\nomenclature[O]{\(w_i/\epsilon\)}{Scalar weight/design parameter.}
\nomenclature[O]{\(P^{\max}/Q^{\max}\)}{Active/reactive maximum line power flow.}
\nomenclature[O]{\(V_{\min}^{\mathtt{sq}}/V_{\max}^{\mathtt{sq}}\)}{Min/max squared voltage magnitude limit.}
\nomenclature[O]{\(\mathbf{y}^{\mathtt{min}}/\mathbf{y}^{\mathtt{max}}\)}{Min/max power exchange limit.}
\nomenclature[O]{\(E^{\mathtt{rated}}\)}{Rated energy capacity of the BESS.}
\nomenclature[O]{\(\Omega^c\)}{Capital cost of the BESS installation.}
\nomenclature[O]{\(c_j\)}{Approximate marginal aging cost.}

\nomenclature[N]{\(\VV\)}{Set of nodes/buses in the distribution network.}
\nomenclature[N]{\(\EE\)}{Set of lines/branches in the distribution network.}
\nomenclature[N]{\(\mathcal{C}_j\)}{Set of child nodes of ADN node $j$.}
\nomenclature[N]{\(\mathcal{P}\)}{Set of prosumers.}
\nomenclature[N]{\(\)}{}
\nomenclature[N]{\(\)}{}
\nomenclature[N]{\(\)}{}
\nomenclature[N]{\(\)}{}
\nomenclature[N]{\(\)}{}

\printnomenclature

\section{Introduction}
\label{sec: Int}

The integration of distributed energy resources (DERs) at the prosumer level within the active distribution network (ADN) has been accelerating in recent years. This transition has enabled prosumers to derive economic benefits by participating in local energy markets, engaging in demand-side management programs, and providing grid services \cite{liu2021grid}. However, it has also introduced challenges for system operators; if the power exports and imports by the prosumers are not properly regulated, it could cause problems such as over-voltage and line congestion, posing threats to the stability of the network. 

\subsection{Related work}
To address these challenges, various solutions have been investigated, including coordinated control of voltage regulation devices such as PV inverters and capacitor banks \cite{yao2021coordinated,othman2019coordinated}, Volt/VAR droop control strategies \cite{singh2020multistage}, and optimal sizing of PV inverters during the planning stage \cite{ali2020probabilistic}. Another widely adopted approach is for system operators to impose limits on the amount of power that prosumers can import from or export to the distribution network. For example, in Australia, several distribution companies have implemented a static export limit of $5$ kW for prosumer power exchange \cite{liu2021grid}. While this limit ensures safe operation of the network, it also reduces economic benefits for prosumers, especially when they have surplus power exceeding the fixed limit. To overcome this limitation, system operators have introduced {\it dynamic operating envelopes} (DOEs) -- dynamically varying export and import limits on the prosumers active and reactive power injection to the ADN. These nodal limits ensure that the physical and operational constraints of the network are not exceeded, and DNOs are not required to actively control behind-the-meter assets of the prosumers \cite{liu2021grid, alam2023allocation,hoque2023dynamic,azim2023dynamic, liu2022using, attarha2021network}. Therefore, among different proposed approaches for facilitating DER integration in the distribution network, DOEs have emerged as a popular choice.

For informed and reliable DOE estimation, it is essential to consider the local decision-making process of prosumers to capture their intended power exchanges with the distribution network. Prosumers typically have local renewable energy generation (such as, rooftop solar PV) capabilities and battery energy storage systems (BESS) available with them, and aim to minimize the cost of energy procurement from the grid and the degradation cost of the BESS unit. However, most of the prior works in the context of DOE computation have only focused on the energy procurement cost.  

Specifically, \cite{petrou2021ensuring} employed a multi-period rolling horizon mixed-integer linear programming (MILP) optimization approach to determine these intended exchanges; it did not include battery degradation costs or the reactive power capabilities of DERs. Another DOE estimation approach proposed in \cite{liu2022using} leveraged on-load tap changers (OLTCs) to manage voltage and expand DOE limits. Nevertheless, it assumed that all prosumers either export or import power simultaneously within a given interval, which is not realistic. A state estimation-based technique was introduced in \cite{alam2023allocation} to calculate DOEs at the transformer connection point between medium-voltage and low-voltage networks. Authors in \cite{kumarawadu2025smart} have utilized smart-meter data to calculate DOE based on voltage sensitivity coefficients. This approach enabled fair distribution of the curtailment among the DER injections. In \cite{mahmoodi2023capacity}, the authors proposed a right-hand side decomposition method to compute DOEs that assist prosumers in estimating the capacity of their DER installations, and also incorporated uncertainty in both load and generation. To address uncertainty in DOE computation, the authors in \cite{yi2022fair} proposed a chance constrained optimal power flow (OPF) approach. More recently, studies such as \cite{liu2024robust,liu2023linear} introduced nodal operating envelopes for prosumers by robustifying the feasibility set of a linearized power flow model. In \cite{lankeshwara2023time} and \cite{lankeshwara2025development}, the authors required each prosumer to report its range of feasible active and reactive power exchange, and then proposed a sampling based approach to compute a polytope which resides in the hyper-rectangle reported by each prosumer such that any power exchange within the respective polytopes satisfy the power-flow constraints. While the authors in \cite{gerdroodbari2022dynamic} included reactive power control in the calculation of operating envelopes, the formulation was limited to PV inverters; storage inverters were not considered. 

None of the above works have incorporated battery degradation cost in the local decision-making process of the prosumers. To the best of our knowledge, \cite{wickramasinghe2025multi} is the only prior work that incorporates electric vehicle (EV) battery aging costs into the operational objective of the EV scheduling problem. However, that study computes DOE limits only for active power exchange, assumes zero reactive power DOE, does not account for prosumer-level local decision making, relying instead on DNO-side load demand forecasts, and overlooks the reactive power capabilities of DERs.

\subsection{Research gap and contributions}

In many of the above reported works, prosumers are required to report their desired power exchange with the DNO. The DNO then checks whether the reported quantities satisfy the line flow and voltage magnitude limits, and if not, reduces the power exchange limit of the prosumers so as to satisfy the constraints. These techniques do not allow the prosumer to exchange a larger quantity of power compared to the reported value. Moreover, when prosumers encounter uncertainty in PV generation and load, and are unable to absorb them in the storage device, the above computed DOE limits can be restrictive and lead to curtailment in PV generation and load. Consequently, it could lead to prosumers strategically inflating their reported power exchange values to the DNO, which is not desirable. Therefore, it is essential to compute DOE limits that also allow for larger amount of power to be exchanged with the grid compared to the reported values when permissible. Furthermore, most existing studies do not incorporate degradation cost of the BESS in the local decision-making problem of the prosumers.

To address the above identified gaps, this work adopts the flexible optimization framework recently developed in \cite{simonetto2024flexible} to compute DOE limits and integrates battery degradation costs into the local decision-making problem of the prosumers. The main contributions of this work are summarized below.
\begin{itemize}   
    \item Prosumer-centric flexible DOEs are computed in a principled manner using a flexible optimization framework that trades off optimality and flexibility. These DOEs establish upper and lower limits on the active and reactive power exchange between each prosumer and the ADN, ensuring that operational constraints of the network are satisfied. A key advantage of this framework is that, by providing distinct upper and lower bounds, it accommodates the potential for prosumers to switch between consumption and generation modes within an interval.
    \item A multi-stage optimization problem is developed to determine the desired active power exchange by each prosumer with the ADN, which is then used to compute the DOEs. The proposed problem incorporates the degradation cost of the BESS.  
\end{itemize}
The proposed framework is validated on a modified Australian low-voltage distribution network. A comparative assessment between the proposed flexible DOE and a non-flexible DOE approach is carried out under uncertainties in prosumer load and generation, with a focus on operational costs, PV and load curtailment and BESS degradation. The results show that the proposed approach lowers overall operating costs primarily through significant reductions in curtailment, and results in a reduction in battery degradation cost. Furthermore, a sensitivity analysis of the design parameters is performed, and the scalability of the framework is examined on larger distribution networks, confirming its suitability for real-time application in larger-scale systems.

The rest of the paper is organized as follows. Section~\ref{sec: Prob} formulates the optimization problem for the prosumers, and presents the power flow model for the ADN. Section~\ref{sec: FOM} introduces the computation of proposed flexible DOEs through a flexible optimization framework. Section~\ref{sec: Num_Sim} presents simulation results and case studies. Section~\ref{Sum_Con} concludes the paper with a brief discussion on directions for future research.


\section{Problem Formulation}
\label{sec: Prob}


\begin{figure}
    \centering
    \includegraphics[width=0.9\linewidth]{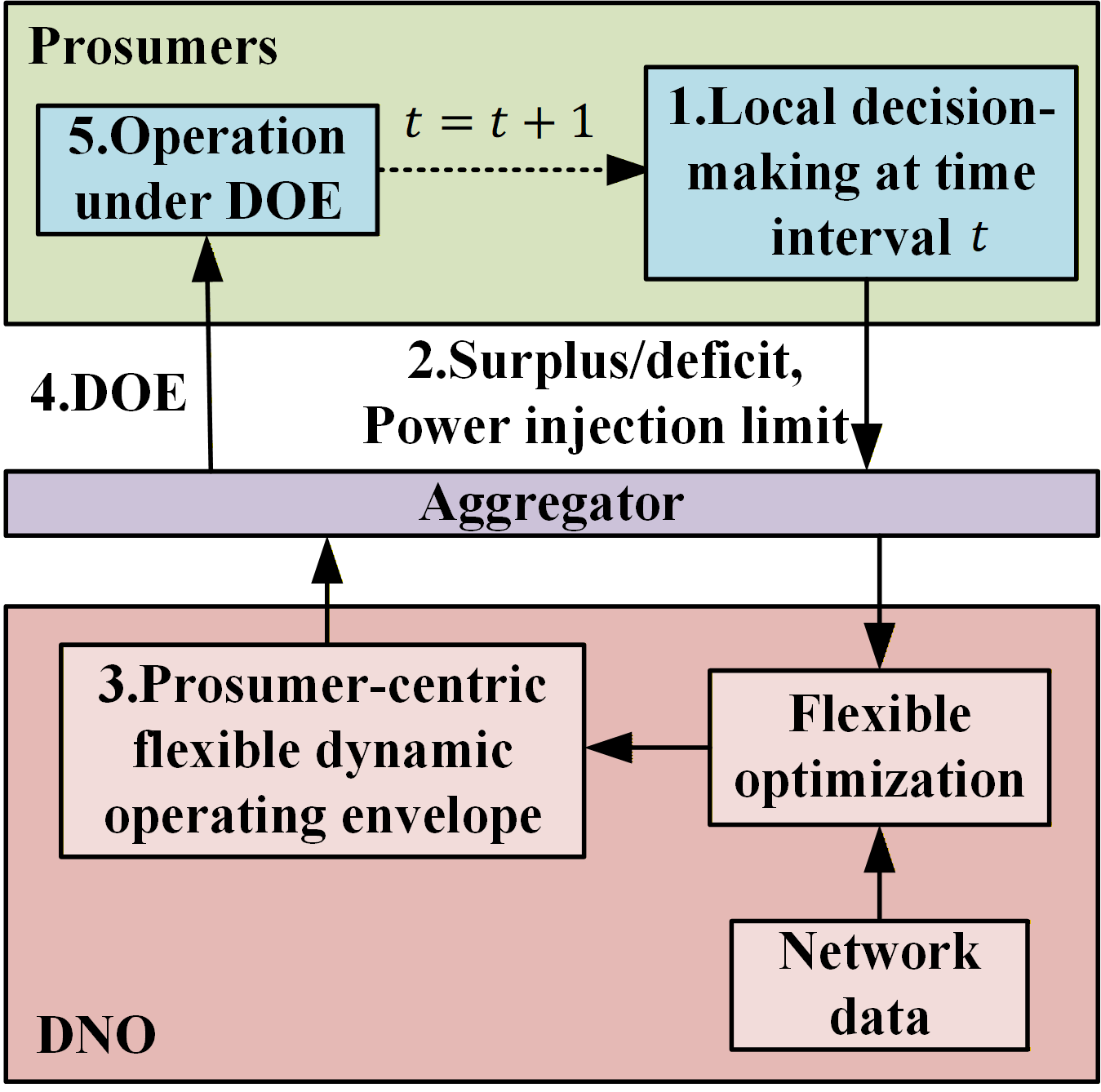} 
    \caption{Hierarchical structure of the proposed system model, illustrating the complete process from prosumers' local decision-making to DOE computation by the DNO, followed by prosumer re-optimization at time interval $t$.}
    \label{SD}
\end{figure}

We consider a low-voltage (LV) ADN comprising of a group of prosumers. The ADN is managed by a DNO which is responsible for safe and reliable operation of the network. Additionally, an aggregator acts as an intermediary between the prosumers and the DNO within the ADN. The proposed system model, shown in Fig. \ref{SD}, illustrates the interaction among prosumers, the aggregator and the DNO. 


Ahead of each interval, prosumers solve an optimization problem locally to determine their desired active power exchange that minimizes energy purchase and battery degradation costs using PV generation and load forecast. Then, they communicate their expected surplus or deficit, along with their maximum power injection capabilities, to the DNO via the aggregator. Based on this information, the DNO estimates prosumer-centric flexible DOE limits for both active and reactive power exchange. These DOE limits are then communicated back to the prosumers. After receiving the DOE limits, the actual PV and load profiles are realized, and each prosumer re-solves its local optimization problem to determine the final operating point within the prescribed DOE limits. This process ensures that the prosumer operation remains feasible when actual PV generation and load differs from their forecast. The complete problem formulation is detailed below.

\subsection{Prosumer optimization problem}
\label{subsec: pros_opt}

\begin{figure}
   \centering
   \includegraphics[width=0.9\linewidth]{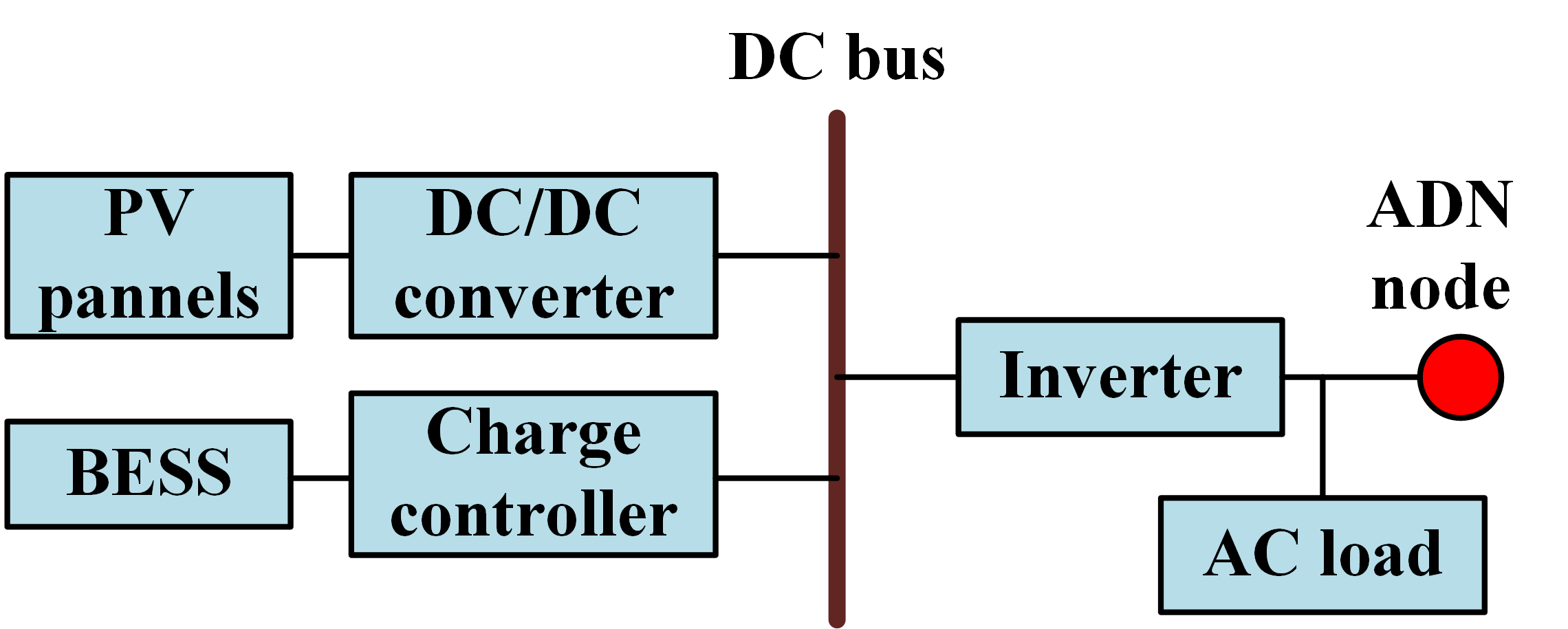} 
   \caption{Internal configuration of a prosumer.}
   \label{Pros}
\end{figure}

Let $\mathcal{P} = \{1,2,\ldots,n\}$ denote the set of $n$ prosumers that are embedded in the ADN. We assume that each prosumer is equipped with local PV generation, a BESS, local load and a controllable inverter which resides at the interface of the prosumer and the ADN as modeled in \cite{zeraati2016distributed}. The PV panels are connected to the DC bus via a DC/DC converter, while the BESS is linked to the DC bus through a charge controller (Fig. \ref{Pros}). The prosumers’ battery can be charged from both the PV system and the grid. We first discuss the battery degradation model considered in this work, followed by formulating the optimization problem that is solved in the first stage. 

\subsubsection{BESS degradation model}
\label{subsubsec: bess_deg}
In this work, we focus on modeling the cyclic aging of the BESS, since calendar aging is negligible over short-term operational horizons \cite{padmanabhan2019battery}. Since the daily charged and discharged energy volumes are approximately equal, we assume that cyclic aging is incurred only during the discharge half-cycle, while charging does not contribute to aging. The cycle depth $\theta_t$ of the BESS at time $t$ is given by \cite{xu2017factoring}:
\begin{align}
    \theta_t=\frac{ \tau P^{\mathtt{dis}}_t}{\eta_{\mathtt{dis}}E^{\mathtt{rated}}}  +\theta_{t-1},
\end{align}
where $P^{\mathtt{dis}}_t$ denotes the discharge power at time $t$, $\tau$ is the sampling interval, $E^{\mathtt{rated}}$ is the rated energy capacity of the BESS and $\eta_{\mathtt{dis}}$ represents its discharge efficiency. Following \cite{xu2017factoring}, the corresponding cycle depth stress function for a lithium-ion battery is modeled as
\begin{align}\label{stress}
    \gamma_t(\theta_t)=5.24 \times 10^{-4}\theta_t^{2.03}.
\end{align}
The marginal aging cost is then defined as-
\begin{align}\label{marg_cost}
   \Omega^c  \frac{\partial \gamma_t(\theta_t)}{\partial P^{\mathtt{dis}}_t}=  \frac{\Omega^c }{\eta_{\mathtt{dis}}E^{\mathtt{rated}}}\frac{d \gamma_t(\theta_t)}{d \theta_t},
\end{align}
where, $\Omega^c$ denotes the capital cost of the BESS installation. Since, \eqref{marg_cost} is nonlinear, directly incorporating it into the prosumer scheduling objective can significantly increase computational complexity. Hence, consistent with prior studies \cite{xu2017factoring}, we employ a piecewise linear upper approximation of \eqref{marg_cost} to model the cyclic life loss of BESS. Considering $J$ linear segments, the approximate marginal aging cost for each segment $j \in \{1,2,\ldots,J\}$ is given by
\begin{align}\label{eq:deg_cost}
c_{j} = \frac{\Omega^c J}{\eta_{\mathtt{dis}} E^{\mathtt{rated}}} \left[ \gamma_t\!\left(\frac{j}{J}\right) - \gamma_t\!\left(\frac{j-1}{J}\right) \right].
\end{align}

\subsubsection{Prosumer desired power exchange computation}
\label{subsubsec: pros_des}

Each prosumer solves a multi-stage constrained optimization problem over a prediction horizon $N_p$ in a receding horizon manner to minimize the total net cost of energy procurement from the grid and cost of using the BESS over the prediction horizon subject to operational constraints.  

We define the decision vector for each prosumer $i \in \mathcal{P}$ at time interval $t$ as
\begin{align*}
    \mathbf{u}_{i,t} := & (P^{\mathtt{ch}}_{i,t}, P^{\mathtt{dis}}_{i,t}, p^{\mathtt{ch}}_{i,j,t}, p^{\mathtt{dis}}_{i,j,t}, e_{i,j,t+1}, \alpha_{i,t}, \pinj_{i,t}, P^{g}_{i,b,t}, P^{g}_{i,s,t}, \\
    & \delta_{i,t}),
\end{align*}
where, $P^{\mathtt{ch}}_{i,t}$ and $P^{\mathtt{dis}}_{i,t}$ represent the charging and discharging powers of the BESS, respectively. To capture battery degradation, a piecewise linear cost model (as given in \eqref{eq:deg_cost}) with $J$ segments is employed. Accordingly, $p^{\mathtt{ch}}_{i,j,t} \in \mathbb{R}^{J}$ and $p^{\mathtt{dis}}_{i,j,t} \in \mathbb{R}^{J}$ represent the charging and discharging powers corresponding to the $j^{\text{th}}$ linear segment, while $e_{i,j,t+1} \in \mathbb{R}^{J}$ denotes the energy stored in that segment at time $(t+1)$. The binary variable $\alpha_{i,t} \in \{0,1\}$ indicates the operating mode of the BESS for prosumer $i$, where $\alpha_{i,t}=1$ corresponds to discharging and $\alpha_{i,t}=0$ to charging. Additionally, let $\pinj_{i,t}$ denotes the active power exchanged with the ADN at time interval $t$, with positive value indicating power export. While, $P^{g}_{i,b,t}$ and $P^{g}_{i,s,t}$ represent the active power purchased from and sold to the grid, respectively. Finally, the binary variable $\delta_{i,t} \in \{0,1\}$ captures the direction of power exchange with the grid, where $\delta_{i,t}=0$ denotes power import and $\delta_{i,t}=1$ denotes power export.

The optimization problem for prosumer $i$ at $t=1$ is:
\begin{subequations}\label{eq:prosumer_optimization}
    \begin{align}
        \min_{\mathbf{u}_{i,1}, \ldots, \mathbf{u}_{i,N_p}} \: \: &\sum_{t=1}^{N_p}  \bigg [ \pi ^{\mathtt{ToU}}_t P^g_{i,b,t}\tau - \pi ^{\mathtt{FiT}}_t P^g_{i,s,t}\tau + \sum_{j=1}^{J} \tau c_{j}p^{\mathtt{dis}}_{i,j,t}\bigg ] ,  \label{eq:prosumer_con_cost} \\
        \text{s.t.} \: \: & e_{i,j,t+1}= e_{i,j,t} + \tau \left(\eta_i^{\mathtt{ch}} p^{\mathtt{ch}}_{i,j,t}-\frac{p^{\mathtt{dis}}_{i,j,t}}{\eta_i^{\mathtt{dis}}}\right), \label{eq:pros_energy_1}\\
        & e_{i,j,t+1} \leq \bar{e}_{i,j}, \label{eq:pros_energy_2}  \\
        & E_{\min,i} \leq \sum_{j=1}^{J}e_{i,j,t+1} \leq E_{\max,i} \label{eq:prosumer_elimit}, \\
        & 0 \leq p^{\mathtt{ch}}_{i,j,t},p^{\mathtt{dis}}_{i,j,t},  \label{eq:pros_power_1}  \\
        & P^{\mathtt{ch}}_{i,t} = \sum_{j=1}^{J} p^{\mathtt{ch}}_{i,j,t}, \label{eq:pros_power_2} \\
        & P^{\mathtt{dis}}_{i,t} = \sum_{j=1}^{J} p^{\mathtt{dis}}_{i,j,t}, \label{eq:pros_power_3} \\
        & 0 \leq P^{\mathtt{dis}}_{i,t} \leq \alpha_{i,t} P^{\mathtt{bat}}_{\max,i} \label{eq:prosumer_crate_dis}, \\
        & 0 \leq P^{\mathtt{ch}}_{i,t} \leq (1-\alpha_{i,t}) P^{\mathtt{bat}}_{\max,i} \label{eq:prosumer_crate_ch}, \\
        & P^{\mathtt{inj}}_{i,t} =  P^{\mathtt{PV},\mathtt{frc}}_{i,t} + P^{\mathtt{dis}}_{i,t}-P^{\mathtt{ch}}_{i,t}-P^{\mathtt{load},\mathtt{frc}}_{i,t}     \label{PB_1}, \\
        & P^{\mathtt{inj}}_{i,t} = P^{g}_{i,s,t} - P^{g}_{i,b,t}    \label{PB_2}, \\
        & P^{\mathtt{inj}}_{\min,i} \leq P^{\mathtt{inj}}_{i,t} \leq P^{\mathtt{inj}}_{\max,i}     \label{Pinj_lim}, \\
        & 0 \leq P^{g}_{i,s,t} \leq \delta_{i,t} P^{\mathtt{inj}}_{\max,i}     \label{PS_lim}, \\
        & 0 \leq P^{g}_{i,b,t} \leq -P^{\mathtt{inj}}_{\min,i}(1 - \delta_{i,t})        \label{PB_lim},
    \end{align}
\end{subequations}
where, the constraints hold for $t \in \{1, 2, \ldots, N_p\}$.

The coefficients ($\pi ^{\mathtt{ToU}}_t, \pi ^{\mathtt{FiT}}_t$) in the cost function \eqref{eq:prosumer_con_cost} correspond to the time-of-use tariff and feed-in-tariff, respectively. The parameters $\eta_i^{\mathtt{ch}}$ and $\eta_i^{\mathtt{dis}}$ denote the charging and discharging efficiencies of the BESS, while $\tau $ is the optimization sampling interval. The energy capacity of the BESS is constrained between $E_{\min,i}$ and $E_{\max,i}$, and the maximum energy associated with the $j^{\text{th}}$ segment of the piecewise linear degradation model is limited by $\bar{e}_{i,j}$. The maximum charging and discharging power of the BESS is bounded by $P^{\mathtt{bat}}_{\max,i}$.

Further, $P^{\mathtt{PV},\mathtt{frc}}_{i,t}$ and $P^{\mathtt{load},\mathtt{frc}}_{i,t}$ denote PV generation and active load demand forecast of prosumer $i\in\mathcal{P}$ for time $t$. The allowable range for active power injection into the ADN bus is bounded by $P^{\mathtt{inj}}_{\min,i}$ and $P^{\mathtt{inj}}_{\max,i}$.

In the objective function \eqref{eq:prosumer_con_cost}, the first two terms capture the cost of energy purchase and the revenue from energy sales, respectively, while the third term accounts for BESS degradation using the piecewise linear model described in Section~\ref{subsubsec: bess_deg}. Constraint \eqref{eq:pros_energy_1} models the dynamic behavior of the BESS, and constraint \eqref{eq:pros_energy_2} limits the maximum energy stored in each degradation segment. Constraint \eqref{eq:prosumer_elimit} enforces overall BESS energy bounds. Constraints \eqref{eq:pros_power_1}–\eqref{eq:pros_power_3} ensure non-negativity and consistency between segment-wise and total charging and discharging powers. The maximum discharging and charging rates are enforced by \eqref{eq:prosumer_crate_dis} and \eqref{eq:prosumer_crate_ch}, while the binary variable $\alpha_{i,t}$ prevents the simultaneous charging and discharging. 

Constraints \eqref{PB_1}–\eqref{PB_2} enforce local power balance at the prosumer level, ensuring that surplus or deficit power is exchanged with the ADN and correctly mapped to power sold or purchased. Constraint \eqref{Pinj_lim} bounds the permissible active power exchange with the ADN, while \eqref{PS_lim} and \eqref{PB_lim} limit the maximum power sold to and purchased from the grid, respectively. The binary variable $\delta_{i,t}$ ensures that power cannot be purchased and sold simultaneously.

The optimization problem defined in \eqref{eq:prosumer_optimization} is a MILP. We denote its optimal solution by $\mathbf{u}_{i,t}^{\mathtt{\star}}$, with the corresponding desired active power exchange between prosumer $i \in \mathcal{P}$ and the ADN represented as $\mathbf{y}^{\mathtt{ref}}_{i,t} := P^{\mathtt{inj},\mathtt{\star}}_{i,t}$. Each prosumer communicates their desired exchange $\mathbf{y}^{\mathtt{ref}}_{i,t}$ with the DNO through aggregator. Based on this information, the DNO determines the prosumer-centric flexible DOEs, as detailed in Section~\ref{sec: FOM}. A key advantage of the flexible DOE framework is that, by providing both upper and lower bounds on active and reactive power exchange, prosumers are not restricted to a fixed buyer or seller role within an interval and can switch their operating mode as needed. After computing the DOE limits, the DNO communicates them back to the prosumers. 

\subsection{Prosumer operation under DOE constraints}
\label{subsec: pros_doe}

During actual operation, the PV generation and load demand differ from their forecasted values. These actual realized values are denoted by $P^{\mathtt{PV},\mathtt{unc}}_{i,t},P^{\mathtt{load},\mathtt{unc}}_{i,t}$ and $Q^{\mathtt{load},\mathtt{unc}}_{i,t}$. For each time interval $t$, the DNO provides flexible DOE limits that are valid only for the current interval. Consequently, prosumers are required to operate strictly within these bounds, curtailing PV generation or load demand if required. Since no future DOE information is available beyond the current interval, each prosumer formulates and solves a single-stage optimization problem at time $t$. This problem minimizes real-time operating and curtailment costs while satisfying the DOE bounds for the current interval, local power balance equations, and DER operational constraints.  

Let us define the decision vector for each prosumer $i \in \mathcal{P}$ at time interval $t$ as 
\begin{align*}
    \bar{\mathbf{u}}_{i,t} := & \big(P^{\mathtt{ch}}_{i,t},P^{\mathtt{dis}}_{i,t}, p^{\mathtt{ch}}_{i,j,t},p^{\mathtt{dis}}_{i,j,t}, e_{i,j,t+1}, \alpha_{i,t}, \pinj_{i,t}, \qinj_{i,t}, \\ 
    & P^{g}_{i,b,t}, Q^{g}_{i,b,t}, P^{g}_{i,s,t}, Q^{g}_{i,s,t}, \delta_{i,t}, P^{\mathtt{inv}}_{i,t}, Q^{\mathtt{inv}}_{i,t}, P^{\mathtt{PV}}_{\mathtt{curt},i,t}, \\
    & P^{\mathtt{load}}_{\mathtt{curt},i,t}\big),
\end{align*}
where, $\qinj_{i,t}$ denotes the reactive power exchanged with the ADN at time interval $t$. The variables $Q^{g}_{i,b,t}$ and $Q^{g}_{i,s,t}$ represent the reactive power purchased from and supplied to the grid, respectively. The inverter's active and reactive power outputs are denoted by $P^{\mathtt{inv}}_{i,t}$ and $Q^{\mathtt{inv}}_{i,t}$. Furthermore, $P^{\mathtt{PV}}_{\mathtt{curt},i,t}$ and $P^{\mathtt{load}}_{\mathtt{curt},i,t}$ denote the curtailed PV generation and load demand, respectively. The remaining variables were already defined in the previous subsection. The active and reactive power DOE limits are given by ($P^{\mathtt{DOE},-}_{i,t}, P^{\mathtt{DOE},+}_{i,t} $) and ($Q^{\mathtt{DOE},-}_{i,t}, Q^{\mathtt{DOE},+}_{i,t}$), respectively. The per-unit curtailment cost coefficients for PV generation and load are denoted by $c^{\mathtt{PV}}_t$ and $c^{\mathtt{load}}_t$.

Each prosumer then solves the following single-stage optimization problem for time interval $t$:
\begin{subequations}\label{eq:pros_opt_post_doe}
    \begin{align}
        \min_{\bar{\mathbf{u}}_{i,t}} \: \: & \big[ \pi ^{\mathtt{ToU}}_t P^g_{i,b,t}\tau - \pi ^{\mathtt{FiT}}_t P^g_{i,s,t}\tau+ \sum_{j=1}^{J} \tau c_{j}p^{\mathtt{dis}}_{i,j,t} + c^{\mathtt{PV}}_tP^{\mathtt{PV}}_{\mathtt{curt},i,t}\tau   \nonumber\\
        & +c^{\mathtt{load}}_tP^{\mathtt{load}}_{\mathtt{curt},i,t}\tau+ c^{\mathtt{bat}} \left(\sum_{j=1}^{J}e_{i,j,t+1}^{\mathtt{\star}}-\sum_{j=1}^{J}e_{i,j,t+1} \right)^2  \big],\label{eq:post_doe_cost} \\
        \text{s.t.} \quad & \eqref{eq:pros_energy_1},\eqref{eq:pros_energy_2},\eqref{eq:prosumer_elimit},\eqref{eq:pros_power_1},\eqref{eq:pros_power_2},\eqref{eq:pros_power_3},\eqref{eq:prosumer_crate_dis},\eqref{eq:prosumer_crate_ch}, \eqref{PB_2}\nonumber \\
        & (P^{\mathtt{inv}}_{i,t})^2 + (Q^{\mathtt{inv}}_{i,t})^2 \leq (S^{\mathtt{inv}}_{i})^2 \label{inv_limit}, \\
        & P^{\mathtt{inv}}_{i,t} = P^{\mathtt{PV},\mathtt{unc}}_{i,t}-P^{\mathtt{PV}}_{\mathtt{curt},i,t} + P^{\mathtt{dis}}_{i,t}-P^{\mathtt{ch}}_{i,t}     \label{PB_1b}, \\
        & P^{\mathtt{inj}}_{i,t} = P^{\mathtt{inv}}_{i,t}-(P^{\mathtt{load},\mathtt{unc}}_{i,t} - P^{\mathtt{load}}_{\mathtt{curt},i,t} )    \label{PB_2b}, \\
        & Q^{\mathtt{inj}}_{i,t} = Q^{\mathtt{inv}}_{i,t}-(Q^{\mathtt{load},\mathtt{unc}}_{i,t} - 0.33P^{\mathtt{load}}_{\mathtt{curt},i,t} )    \label{PB_3b}, \\
        & 0 \leq P^{\mathtt{load}}_{\mathtt{curt},i,t} \leq P^{\mathtt{load},\mathtt{unc}}_{i,t}, \label{Lcurt} \\
        & 0 \leq P^{\mathtt{PV}}_{\mathtt{curt},i,t} \leq P^{\mathtt{PV},\mathtt{unc}}_{i,t}, \label{PVcurt}    \\
        & P^{\mathtt{DOE},-}_{i,t} \leq P^{\mathtt{inj}}_{i,t} \leq P^{\mathtt{DOE},+}_{i,t}    \label{Pinj_limb} , \\
        & Q^{\mathtt{DOE},-}_{i,t} \leq Q^{\mathtt{inj}}_{i,t} \leq Q^{\mathtt{DOE},+}_{i,t}     \label{Qinj_limb}, \\
        & 0 \leq P^{g}_{i,s,t} \leq \delta_{i,t} P^{\mathtt{DOE},+}_{i,t}     \label{PS_limb}, \\
        & 0 \leq P^{g}_{i,b,t} \leq (1 - \delta_{i,t}) |P^{\mathtt{DOE},-}_{i,t}|        \label{PB_limb}.
    \end{align}
\end{subequations}
The quantity $\sum_{j=1}^{J}e_{i,j,t+1}^{\mathtt{\star}}$ represents the total energy stored in the BESS at time ($t+1$), as obtained from the multi-stage optimization problem in \eqref{eq:prosumer_optimization}, $c^{\mathtt{bat}}$ denotes a penalty factor and $S^{\mathtt{inv}}_{i}$ is the apparent power rating of the inverter.

In the objective function \eqref{eq:post_doe_cost}, the first three terms are defined as before. The fourth and fifth terms capture the costs associated with PV and load curtailment, respectively, and the final term penalizes deviations of the BESS energy level from the optimal value obtained in \eqref{eq:prosumer_optimization}. Constraint \eqref{inv_limit} enforces the inverter’s active and reactive power limits. Constraint \eqref{PB_1b} links the inverter active power output to the net contribution of curtailed PV generation and BESS output. Constraints \eqref{PB_2b} and \eqref{PB_3b} define the net active and reactive power exchanged with the ADN, with $0.33$ being the ratio of reactive and active power at $0.95$ power factor. Constraints \eqref{Lcurt} and \eqref{PVcurt} bound the load and PV curtailments, respectively. Compliance with the active and reactive DOE limits is ensured by \eqref{Pinj_limb} and \eqref{Qinj_limb}. Finally, constraints \eqref{PS_limb} and \eqref{PB_limb} limit the active power sold to and purchased from the ADN, while the binary variable $\delta_{i,t}$ prevents simultaneous power import and export.

The inverter constraint in \eqref{inv_limit}, although convex, is nonlinear, which increases the computational burden for the solver. To address this, a piecewise linear approximation is adopted following the approach in \cite{mallick2024distributed}. Let $l$ denote the number of piecewise linear segments used to approximate constraint \eqref{inv_limit}, which enables its reformulation in a linear form as
\begin{align}
    a_j^p P^{\mathtt{inv}}_{i,t} + b_j^q Q^{\mathtt{inv}}_{i,t} + c_{i,j}^s \leq 0, j \in \{1,2,\ldots,l\}.
\end{align}
In this study, $l$ is set to $24$. The linear coefficients in the above expression are derived in \cite{akbari2014linearized} and can be written as
\begin{subequations}
    \begin{align}
        &a_j^p = 2\sin(\pi/l)\sin(\frac{\pi}{l}(2j-1)),\\
        &b_j^q = 2\sin(\pi/l)\cos(\frac{\pi}{l}(2j-1)),\\
        &c_{i,j}^s = -S^{\mathtt{inv}}_{i}\sin(2\pi/l).
    \end{align}
\end{subequations}

The resulting BESS energy level is carried over to initialize the battery state in the next interval. The prosumer optimization formulations remain consistent across intervals, with only the load and generation forecasts being updated. The following section outlines power flow model of the distribution network.

\begin{figure}
   \centering
   \includegraphics[scale= 1.2] {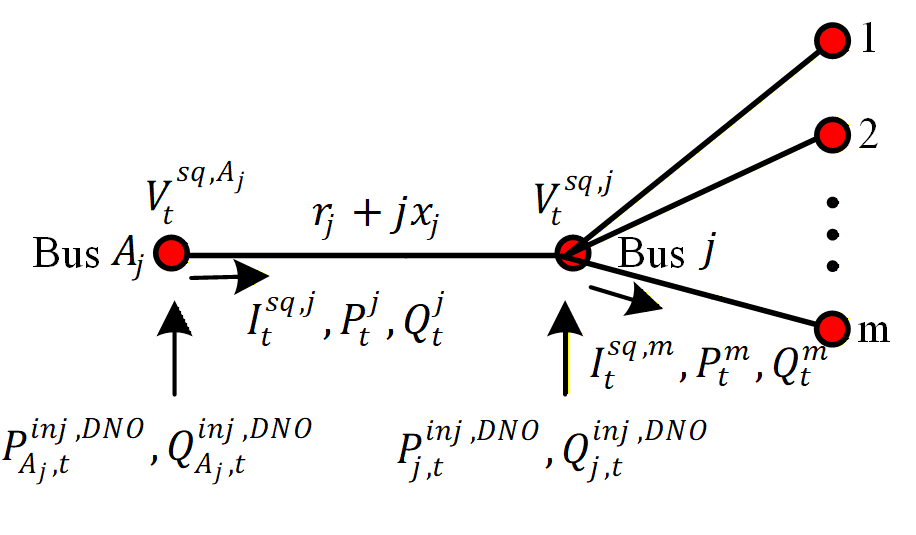}
   \caption{Notation used in the \textit{Distflow} model}
   \label{branchflow_Fig}
\end{figure}


\subsection{Distribution network power flow}

A LV radial distribution network is modeled as a directed acyclic graph $\GG = (\VV,\EE)$ where $\VV$ represents the set of nodes (or buses) and $\EE$ denotes the set of edges (or branches). In this structure, every node $j$, except the substation, has a single ancestor node, denoted by $A_j$, and a collection of child nodes, represented as $\mathcal{C}_j$. For notational convenience, we index the edge $(A_j, j) \in \EE$ which corresponds to the branch associated with node $j$, as line $j$. Each line $j$ is characterized by a resistance $r_j$ and a reactance $x_j$. The substation bus is indexed as node $0$, which, without loss of generality, is assumed to be the point of interconnection between the ADN and the upper voltage level network. The graph is oriented so that each line points away from the substation, or root node. In Fig. \ref{branchflow_Fig}, this orientation is illustrated, with power flow in line $j$ assumed to be directed from its parent node $A_j$ to node $j$. The figure also depicts other relevant quantities, including nodal power injections ($P^{\mathtt{inj},\mathtt{DNO}}_{j,t},Q ^{\mathtt{inj},\mathtt{DNO}}_{j,t}$), square of line currents ($I^{\mathtt{sq},j}_{t}$), and square of bus voltages ($V^{\mathtt{sq},j}_{t}$).

To model power flow within the radial distribution network, we utilize the relaxed branch flow formulation, commonly referred to as the \textit{Distflow} model \cite{peng2016distributed}. As highlighted in \cite{peng2016distributed,mallick2024distributed} this branch flow approach offers improved numerical stability compared to the bus injection model. Let $(P^{\mathtt{inj},\mathtt{DNO}}_{j,t},Q ^{\mathtt{inj},\mathtt{DNO}}_{j,t})_{j \in \VV}$ denote the nodal power injection at bus $j$ at time step $t$. Given the sending-end active ($P^j_{t}$) and reactive ($Q^j_{t}$) power flow in line $j$, the corresponding receiving-end power flow is expressed as:
\begin{subequations}\label{eq:distflow}
\begin{align}
    & \sum_{m\in  \mathcal{C}_j} \: \: (P^m_{t}) = \: P^j_{t} - I^{\mathtt{sq},j}_{t} r_j + P^{\mathtt{inj},\mathtt{DNO}}_{j,t}, \label{eq:distflow_1}
    \\& \sum_{m\in  \mathcal{C}_j} \: \: (Q^m_{t}) = \: Q^j_{t} - I^{\mathtt{sq},j}_{t} x_j + Q^{\mathtt{inj},\mathtt{DNO}}_{j,t}, \label{eq:distflow_2}
    \\& V^{\mathtt{sq},j}_{t} = \: V^{\mathtt{sq},A_j}_{t} - 2(r_jP^j_{t} + x_{j}Q^j_{t}) +  I^{\mathtt{sq},j}_{t}(r^{2}_{j} + x^2_j),  \label{eq:distflow_3}
    \\& I^{\mathtt{sq},j}_{t} =\: \frac{(P^{j}_{t})^2 + (Q^j_{t})^2}{ V^{\mathtt{sq},A_j}_{t}} , \:  \: \label{eq:distflow_4}
\end{align}
\end{subequations}
\sloppy where $V^{\mathtt{sq},A_j}_{t}$ is the squared voltage magnitude at node $A_j$ (ancestor of node $j$), and $I^{\mathtt{sq},j}_{t}$ is the squared current magnitude in line $j$. The constraints \eqref{eq:distflow_1} and \eqref{eq:distflow_2} hold for all $j \in \EE$, while the constraints \eqref{eq:distflow_3} and \eqref{eq:distflow_4} hold for all $j \in \VV$. In particular, $P^1_{t}$ represents the power drawn by the distribution network from the upper-level network.

In the above branch flow model, equations \eqref{eq:distflow_1}-\eqref{eq:distflow_3} are linear functions of the power flow variables $(P^j_{t},Q^j_{t},V^{\mathtt{sq},A_j}_{t},I^{\mathtt{sq},j}_{t})$ while \eqref{eq:distflow_4} is nonlinear. Inspired by past works such as \cite{mallick2024distributed}, we consider a linearization of \eqref{eq:distflow_4} using first order Taylor's series approximation around an initial estimate of $\bar{I}^{\mathtt{sq},j}_{t}$ at $(\bar{P}^j_{t},\bar{Q}^j_{t},\bar{V}^{\mathtt{sq},A_j}_{t})$ given by
\begin{align}
&I^{\mathtt{sq},j}_{t} =  \bar{I}^{\mathtt{sq},j}_{t} + (P^j_{t}-\bar{P}^j_{t})\Big[\frac{\partial I^{\mathtt{sq},j}_{t}}{\partial P^j_{t}}\Big]_{\bar{I}^{\mathtt{sq},j}_{t}} \label{eq:taylor} \\
& +\! (Q^j_{t}\!-\!\bar{Q}^j_{t})\Big[\frac{\partial I^{\mathtt{sq},j}_{t}}{\partial Q^j_{t}}\Big]_{\bar{I}^{\mathtt{sq},j}_{t}} \!\!\!\!+\!\! (V^{\mathtt{sq},A_j}_{t}\!\!-\! \bar{V}^{\mathtt{sq},A_j}_{t})\Big[\frac{\partial I^{\mathtt{sq},j}_{t}}{\partial \bar{V}^{\mathtt{sq},A_j}_{t}}\Big]_{\bar{I}^{\mathtt{sq},j}_{t}}\!\!\!. \nonumber 
\end{align}

The consequence of the above power flow equations is that the power flow on the lines and the voltage magnitude at the nodes can be expressed as a linear function of nodal injection. Let $\mathbf{x}_t=(P^{\mathtt{inj},\mathtt{DNO}}_{j,t},Q^{\mathtt{inj},\mathtt{DNO}}_{j,t})_{j \in \VV}$. A careful observation and rearrangement of the power flow equations \eqref{eq:distflow_1}, \eqref{eq:distflow_2}, \eqref{eq:distflow_3} and \eqref{eq:taylor} leads to elimination of equation \eqref{eq:taylor}. Then, the vector of active and reactive power flows on the lines and the magnitude of nodal voltages can be expressed as
\begin{subequations}\label{power_flow}
\begin{align}
    & P_{t} = A^1_t \mathbf{x}_{t} + b^1_t,
    \\ & Q_{t} = A^2_t \mathbf{x}_{t} + b^2_t,
    \\ & V^{\mathtt{sq}}_{t} = A^3_t \mathbf{x}_{t} + b^3_t,
\end{align}
\end{subequations}
where $A^i_t$ and $b^i_t$, $i\in\{1,2,3\}$, are matrices and vectors of suitable dimensions constructed from \eqref{eq:distflow} and \eqref{eq:taylor}, with
their explicit forms provided in the Appendix. The DNO aims to maintain $ P_{t}, Q_{t}$ and $V^{\mathtt{sq}}_{t}$ within acceptable bounds by determining the DOE at prosumer nodes, thereby indirectly regulating their power exchange with the ADN.

\section{Flexible Optimization Framework}
\label{sec: FOM}

In this section, the proposed flexible DOE computation approach is described, where flexible refers to the ability to define adjustable upper and lower bounds on prosumer power exchange rather than enforcing a single fixed limit. The DNO receives the desired active power exchange from prosumers, obtained after solving \eqref{eq:prosumer_optimization} and denoted by $\mathbf{y}^{\tt ref}_t$, via the aggregator. However, these values may violate network constraints such as voltage constraints or line flow limits. In order to ensure safe operation of the distribution network, the DNO aims to compute upper and lower bounds on the active and reactive power exchange by each prosumer such that prosumers are able to exchange their desired amounts if possible and any power exchange that lie within the DOE bounds satisfy the network constraints. 

To compute these prosumer-centric DOEs, a robust optimization problem is formulated in the flexible optimization framework proposed recently in \cite{simonetto2024flexible}. The decision variables consist of the tuple $(\mathbf{x}_{t},\beta_t) \in \Rb^{4n}$ where the former denotes the nominal active and reactive power injections and $\beta_t\geq 0$ denotes the associated {\it flexibility} margins. In particular, $\mathbf{x}^i_{t} + \beta^i_t$ denotes the upper DOE limit for active (reactive) power exchange if $i \leq n$ ($n+1 \leq i \leq 2n$) and $\mathbf{x}^i_{t} - \beta^i_t$ denotes the corresponding lower DOE limit. 

The main challenge is to ensure that any power exchange from within the DOE limits satisfy network constraints. Note that $(\mathbf{x}_{t} + \text{diag}(\beta_t) \mathbf{z})$ defines a vector of active and reactive power exchange that belongs to the DOE limits whenever $\mathbf{z} \in [-1,1]^{2n}$; here $\text{diag}(\beta_t)$ denotes a $2n \times 2n$ diagonal matrix with the vector $\beta_t$ on its diagonal. Consequently, the DNO solves the following optimization problem at time step $t$: 
\begin{subequations}\label{eq:dno_optimization}
    \begin{align}
        \min_{\mathbf{x}_{t},\beta_t \geq 0} & \quad  \sum^{n}_{i=1} \left(\mathbf{x}_{i,t} - \mathbf{y}^{\mathtt{ref}}_{i,t}\right)^2 + \sum^{2n}_{i=1} w_i \phi(\beta_{i,t}) \label{eq:dno_cost}
        \\ \text{s.t.}  & \quad |A^1_t (\mathbf{x}_{t} + \text{diag}(\beta_t) \mathbf{z}) + b^1_t| \leq P^{\max}, \label{eq:dno_con1}
        \\ & \quad |A^2_t (\mathbf{x}_{t} + \text{diag}(\beta_t) \mathbf{z}) + b^2_t| \leq Q^{\max}, \label{eq:dno_con2}
        \\ & \quad V_{\min}^{\mathtt{sq}} \leq A^3_t (\mathbf{x}_{t} + \text{diag}(\beta_t) \mathbf{z}) + b^3_t \leq V_{\max}^{\mathtt{sq}}, \label{eq:dno_con3}
        \\ & \quad \mathbf{y}^{\mathtt{min}} \leq \mathbf{x}_{t} + \text{diag}(\beta_t) \mathbf{z} \leq \mathbf{y}^{\mathtt{max}},   \label{eq:dno_con4}
    \end{align}
\end{subequations}
where the constraints hold for all $\mathbf{z} \in [-1,1]^{2n}$, and $P^{\max},Q^{\max},V_{\max}^{\mathtt{sq}}$ and $V_{\min}^{\mathtt{sq}}$ are permissible limits for the line flows and squared nodal voltage magnitudes. The vectors $\mathbf{y}^{\mathtt{min}}$ and $\mathbf{y}^{\mathtt{max}}$ denote the minimum and maximum power exchange capability available at the prosumers. 

The cost function \eqref{eq:dno_cost} is convex in the decision variables. The first term in the cost function ensures that the optimal $\mathbf{x}_{i,t}^\star \: \forall i \in \{1,2\dots n\}$ remains close to the desired power injection $\mathbf{y}^{\mathtt{ref}}_{i,t} \: \forall i \in \{1,2\dots n\}$ reported by the prosumers, while the second term determines the optimal flexibility margins $\beta_t^\star$, which allow possible deviations in the prosumer power exchange around $\mathbf{x}_{t}^\star$. The function $\phi(\beta) := -\beta + \frac{\epsilon}{2} \beta^2$ achieves a trade-off between flexibility and optimality \cite{simonetto2024flexible}, governed by the design parameter $\epsilon$. A smaller $\epsilon$ favors larger flexibility at the expense of optimality, and vice versa. The weights $w_i \geq 0$ are prosumer-specific and also reflect the relative importance of flexibility. Note that prosumers do not report any desired reactive power exchange, and hence, there is no term present in the cost function that corresponds to reactive power setpoints. 

Constraints \eqref{eq:dno_con1} and \eqref{eq:dno_con2} ensure that line flows remain within safe limits, whereas, \eqref{eq:dno_con3} maintains bus voltages within permissible bounds for any injection within the DOE bounds. The linear constraints are required to hold in a robust manner for all $\mathbf{z} \in [-1,1]^{2n}$. Unlike the conventional robust and stochastic formulations in \cite{simonetto2024flexible}, we propose a reformulation of the robust constraints exploiting the nature of the constraints. Note that constraints in \eqref{eq:dno_optimization} are of the following form:
\begin{align}\label{eq:robust_reformulation}
    M \text{diag}(\beta_t) \mathbf{z} \leq v, \qquad \forall \mathbf{z} \in [-1,1]^{2n},
\end{align}
where $M$ is a matrix and $v$ is a vector that depends on $\mathbf{x}_t$. The $i^{\text{th}}$ row of the above constraint can be equivalently stated as
\begin{align*}
    \sup_{\mathbf{z} \in [-1,1]^{2n}} \left[M \text{diag}(\beta_t)\right]_i^{\top} \mathbf{z} \leq v_i. 
\end{align*}
It can further be written in the following form-
\begin{align*}
    \sup_{\mathbf{z} \in [-1,1]^{2n}} \sum_{j=1}^{2n} [M]_{ij} \: \beta_{t,j}\: \mathbf{z}_j.
\end{align*}
Consequently, the worst case value is realized when
\[
\mathbf{z}_j^{\mathtt{\star}} =
\begin{cases}
1 & \text{if } [M]_{ij}\geq 0, \\
-1 & \text{if } [M]_{ij}\leq 0, \\
\end{cases}
\]
which yields,
\begin{align*}
    \sup_{\mathbf{z} \in [-1,1]^{2n}} \left[M \text{diag}(\beta_t)\right]_i^{\top} \mathbf{z}=\sum_{j=1}^{2n} |[M]_{ij}| \:\beta_{t,j}.
\end{align*}
Substituting this in the uncertain constraint \eqref{eq:robust_reformulation} replaces it by the following in the optimization problem \eqref{eq:dno_optimization}:
\begin{align*}
|M| \beta_t  \leq v.
\end{align*}
The above constraints are affine in the decision variables $\mathbf{x}_t$ and $\beta_t$, and consequently, \eqref{eq:dno_optimization} remains a convex optimization problem. From the optimal solution of \eqref{eq:dno_optimization}, denoted by $\mathbf{x}^\star_t, \beta^\star_t$, the upper and lower DOE limits on the active and reactive power exchange are given by:
\begin{subequations}\label{DOE_lim}
\begin{align}
&P^{\mathtt{DOE},+}_{i,t}=\mathbf{x}^{\star}_{i,t}+\beta^\star_{i,t}, P^{\mathtt{DOE},-}_{i,t}=\mathbf{x}^{\star}_{i,t}-\beta^\star_{i,t},
\\&Q^{\mathtt{DOE},+}_{i,t}=\mathbf{x}^{\star}_{n+i,t}+\beta^\star_{n+i,t}, Q^{\mathtt{DOE},-}_{i,t}=\mathbf{x}^{\star}_{n+i,t}-\beta^\star_{n+i,t}.
\end{align}
\end{subequations}
Specifically, if prosumers inject power into the ADN such that:
$\pinj_{t} \in [P^{\mathtt{DOE},-}_t,P^{\mathtt{DOE},+}_t]$ and $\qinj_{t} \in [Q^{\mathtt{DOE},-}_t,Q^{\mathtt{DOE},+}_t]$, then, the power flow constraints \eqref{eq:dno_con1}-\eqref{eq:dno_con3} remain satisfied. 

The computed DOE bounds are shared with the prosumers, who incorporate them during real-time operation by adjusting BESS operation or curtailing PV/load as applicable via \eqref{eq:pros_opt_post_doe}. 


\section{Numerical simulation}
\label{sec: Num_Sim}

\begin{figure}
    \centering
    \includegraphics[width=0.85\linewidth]{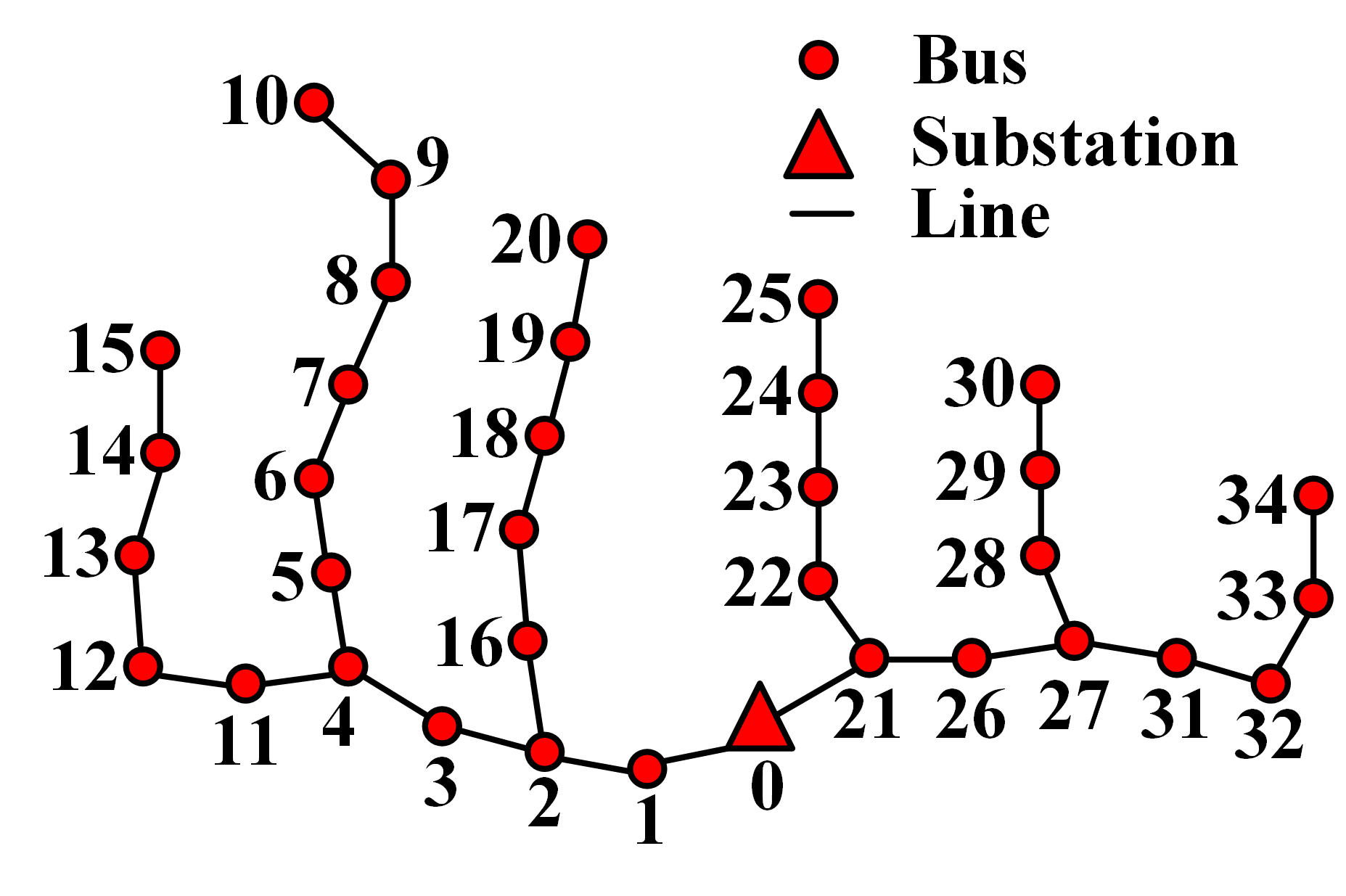}
    \caption{Modified Australian low-voltage distribution network under study \cite{wang2019voltage}.}
    \label{test system}
\end{figure}

\begin{figure}
\centering
   \begin{subfigure}[b]{0.45\textwidth}
            \centering
            \includegraphics[width=0.9\linewidth]{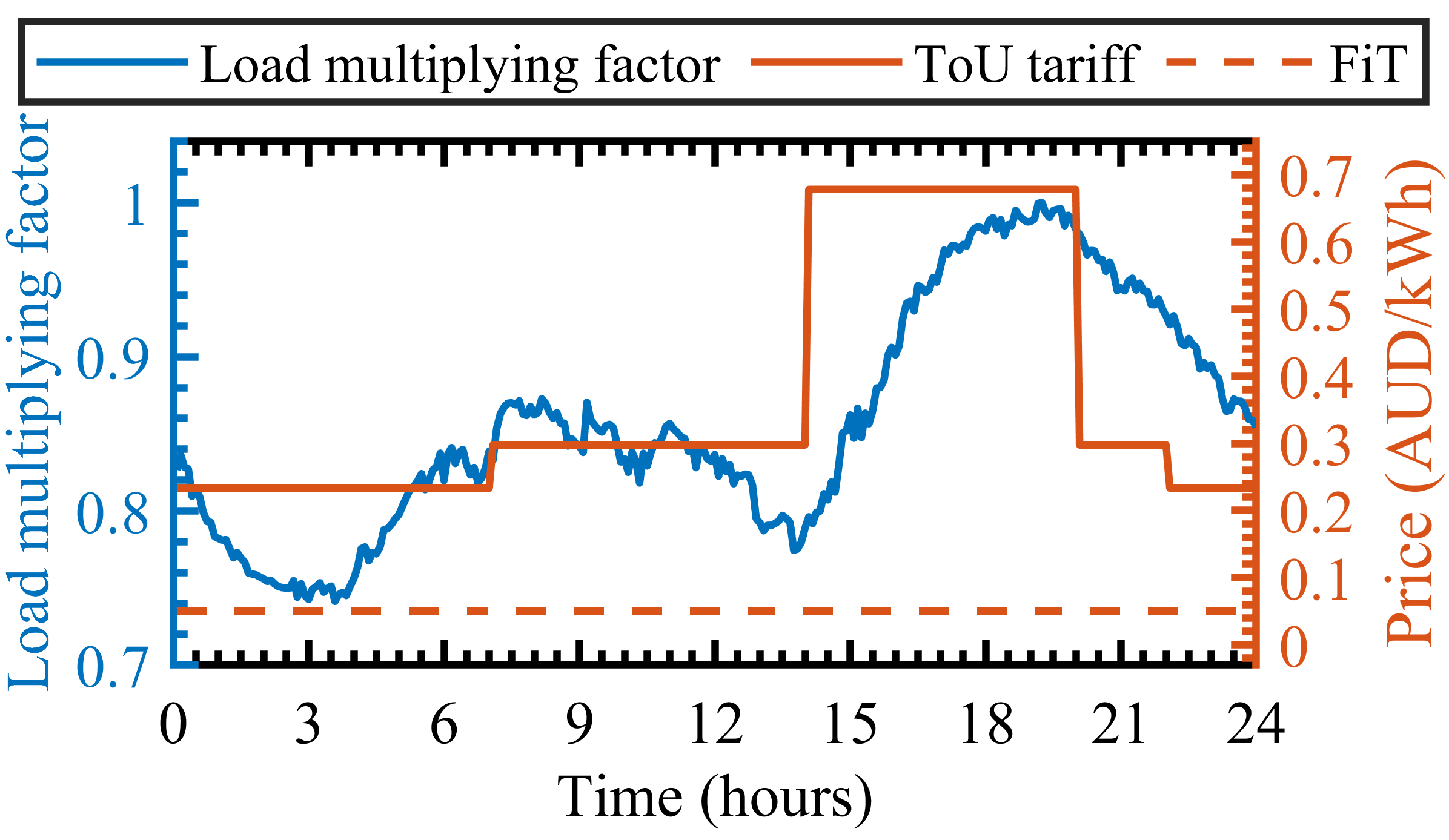}
            \caption{}
            \label{LP}
    \end{subfigure}
    \begin{subfigure}[b]{0.45\textwidth}
            \centering
            \includegraphics[width=0.9\linewidth]{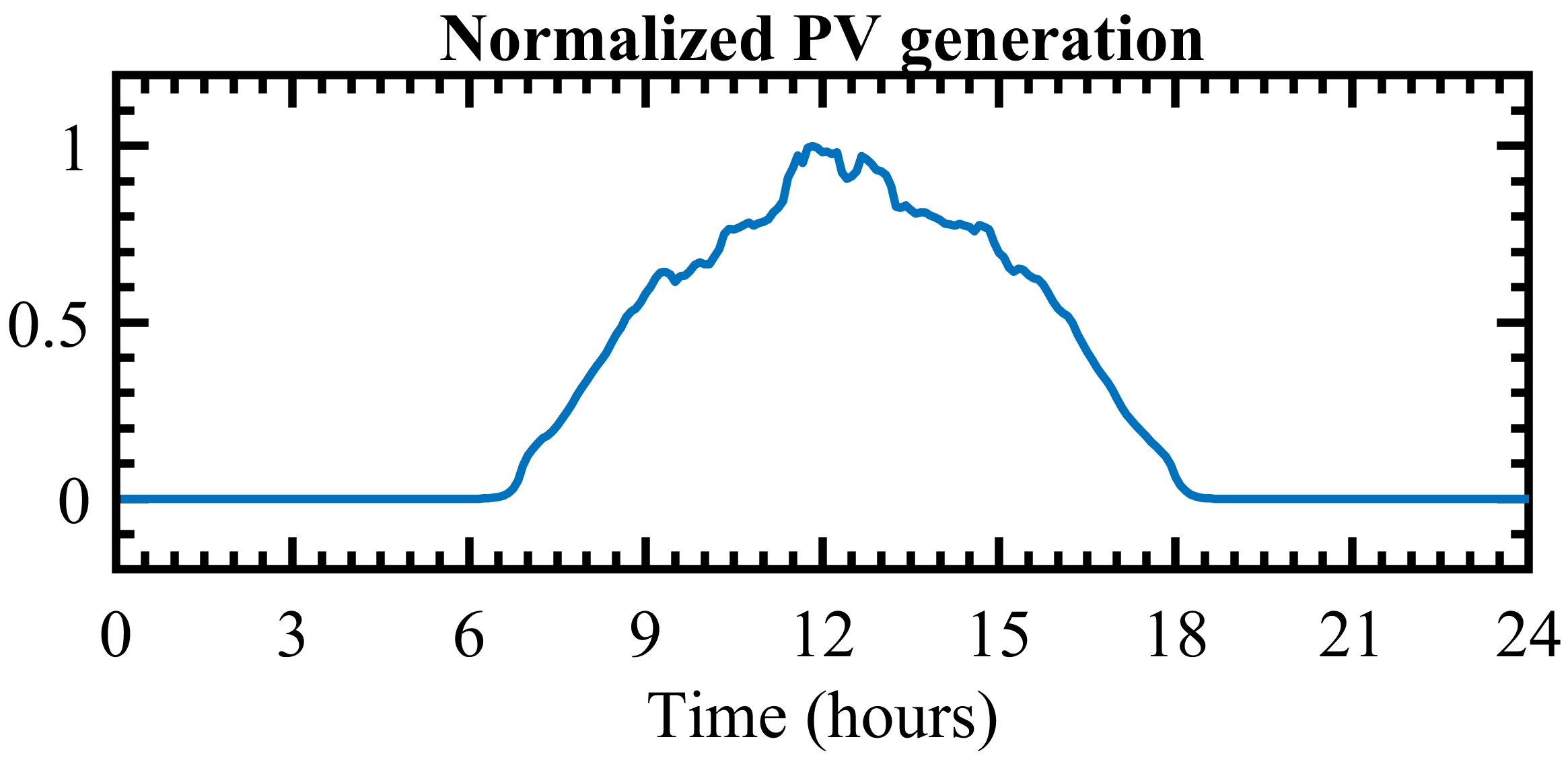} 
            \caption{}
            \label{PV}
    \end{subfigure}
    \caption{(a) Time-of-use tariff, feed-in-tariff and load multiplying factor over a $24$-hour period, (b) normalized PV generation profile of the prosumers.}\label{pv_price}
\end{figure}

The effectiveness of the proposed DOE computation using a flexible optimization approach is demonstrated on a modified Australian low-voltage distribution network \cite{wang2019voltage}, assumed to be balanced, as shown in Fig. \ref{test system}. Since the original model includes only pole-level details, it is extended by assuming one prosumer is connected to each pole. The substation is labeled as bus $0$, while the remaining buses are indexed from $1$ to $34$. Moreover, each of these $34$ buses have one prosumer connected to them.  Each prosumer owns rooftop PV generation, a BESS and an AC load. Additionally, prosumers are equipped with controllable inverters that enable BESS charging from the grid and facilitate power output from both battery discharge and PV generation. Prosumer base active loads range from $2$ to $5.5$ kW and reactive loads are set at $0.33$ times the active load to maintain $0.95$ power factor at the point of connection. Prosumer load profiles vary throughout the day based on a load multiplying factor, as shown in Fig. \ref{LP}. This figure also presents the variation in energy prices, which follows the Alinta Energy Retailer rates in New South Wales (NSW), Australia \cite{energyprice}. The load multiplying factor represents the normalized load variation for January $10$, $2024$, in NSW, Australia, sourced from AEMO \cite{AEMO}. To ensure diversity among prosumer load profiles, a random variation of up to $\pm 2\%$ is applied to the base profile shown in Fig. \ref{LP} and individually assigned to each prosumer.

The PV generation profiles of the prosumers are constructed by scaling their peak PV capacities with a normalized PV generation profile shown in Fig. \ref{PV}. The normalized profile is sourced from NREL and corresponds to a 69 MW solar plant \cite{PVdata}. The peak PV capacities of the prosumers are listed in Table \ref{Tab:PV_bat} and are intentionally selected to be relatively high in order to assess the impact of the proposed approach on PV and load curtailment. 

\begin{table}[h!]
\caption{PV and BESS data}
  \label{Tab:PV_bat}
  \centering
  \footnotesize
  \begin{tabular}{|c|c|c|c|}
    \hline
    Base load                & Peak PV              &  BESS capacity   & Charge/discharge      \\
    
    (kW)                     & generation(kW)       & (kWh)             &  limit (kW)                    \\
    \hline
    Load $\geq 5$            & $9$                     &    $13.1$          &    $4$                       \\
    \hline
    Load $\in [4,5)$      & $8$                     &   $13.1$           &   $4$                        \\
    \hline
     Load $\in [3,4)$     & $7$                     &   $9.89$           &     $3$                        \\
    \hline
    Load $< 3$                & $6$                      &   $6.5$           &  $2$                          \\
    \hline
  \end{tabular}
  \end{table}

The BESS capacity and charge/discharge limit for all prosumers are based on data from LG energy solutions \cite{BESS} and summarized in Table \ref{Tab:PV_bat}. The battery state of charge (SOC), defined as the ratio of stored energy to rated capacity, is maintained within a range of $20\%$ to $80\%$, with an initial SOC of $20\%$ at the beginning of the day. The capital cost of BESS installation is assumed to be $\Omega^c = 300 \: \times$ (maximum BESS capacity in kWh) AUD \cite{mongird2019energy}. Moon conductors are used for the phase conductors in the network, characterized by an AC resistance of $0.284$ ohms/km and the line flow capacity is rated as $100$ kW \cite{nexans}. Consistent with network constraints and installed DER and load capacities, the maximum allowable power injection at prosumer nodes is set to $5.5$ kW for prosumers $1$–$20$ and $6$ kW for prosumers $21$–$34$. 

The simulations are conducted using the MATLAB-based modeling package YALMIP \cite{1393890}, with Gurobi \cite{gurobi} employed as the solver under default settings. Voltage calculations are carried out via load-flow analysis using MATPOWER \cite{matpower}.  All computations are performed on a desktop computer equipped with a $2.90$ GHz Intel Core i$7$ processor and $64$ GB of RAM. Sampling time for the multi-stage optimization problem of the prosumers is set to $5$ minutes. 

\subsection{Impact of the parameter $\epsilon$ on flexible DOE}

\begin{figure*}
    \centering
    \includegraphics[width=0.9\linewidth]{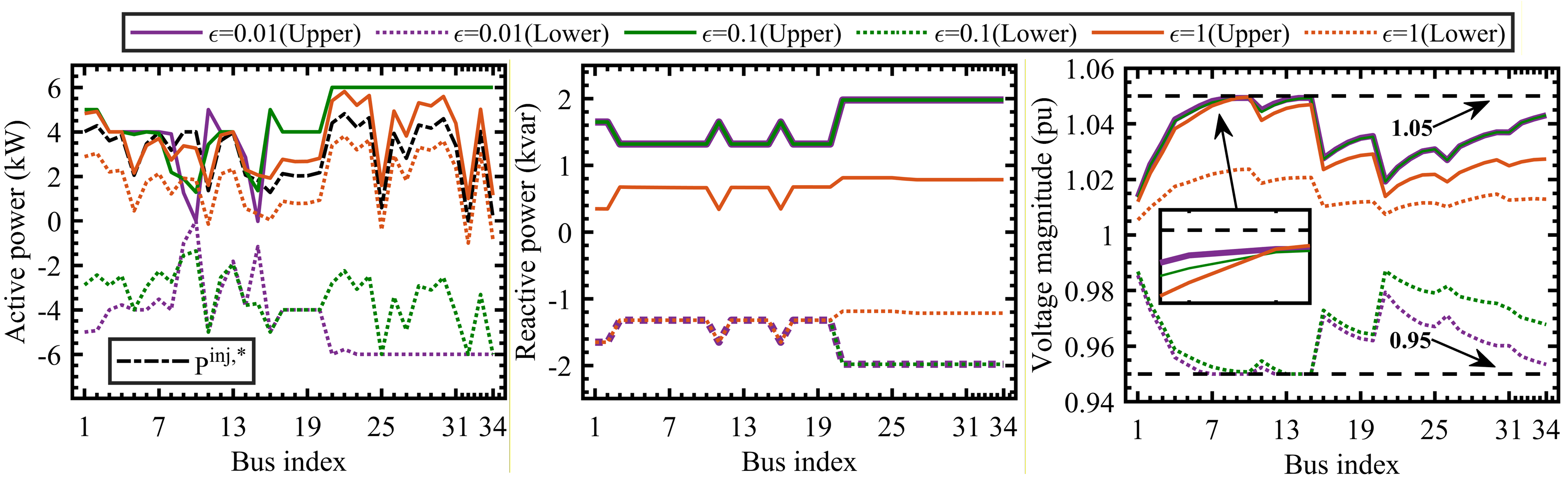} 
    \caption{Sensitivity analysis with varying $\epsilon$ at $12{:}00$: (left–middle) show upper and lower DOE limits for active and reactive power per prosumer, based on preferred setpoint ($P^{\mathtt{inj},\star}$); (right) shows the bus voltage magnitude corresponding to upper and lower DOE limits.}
    \label{eps}
\end{figure*}

\begin{figure}
    \centering
    \includegraphics[width=0.75\linewidth]{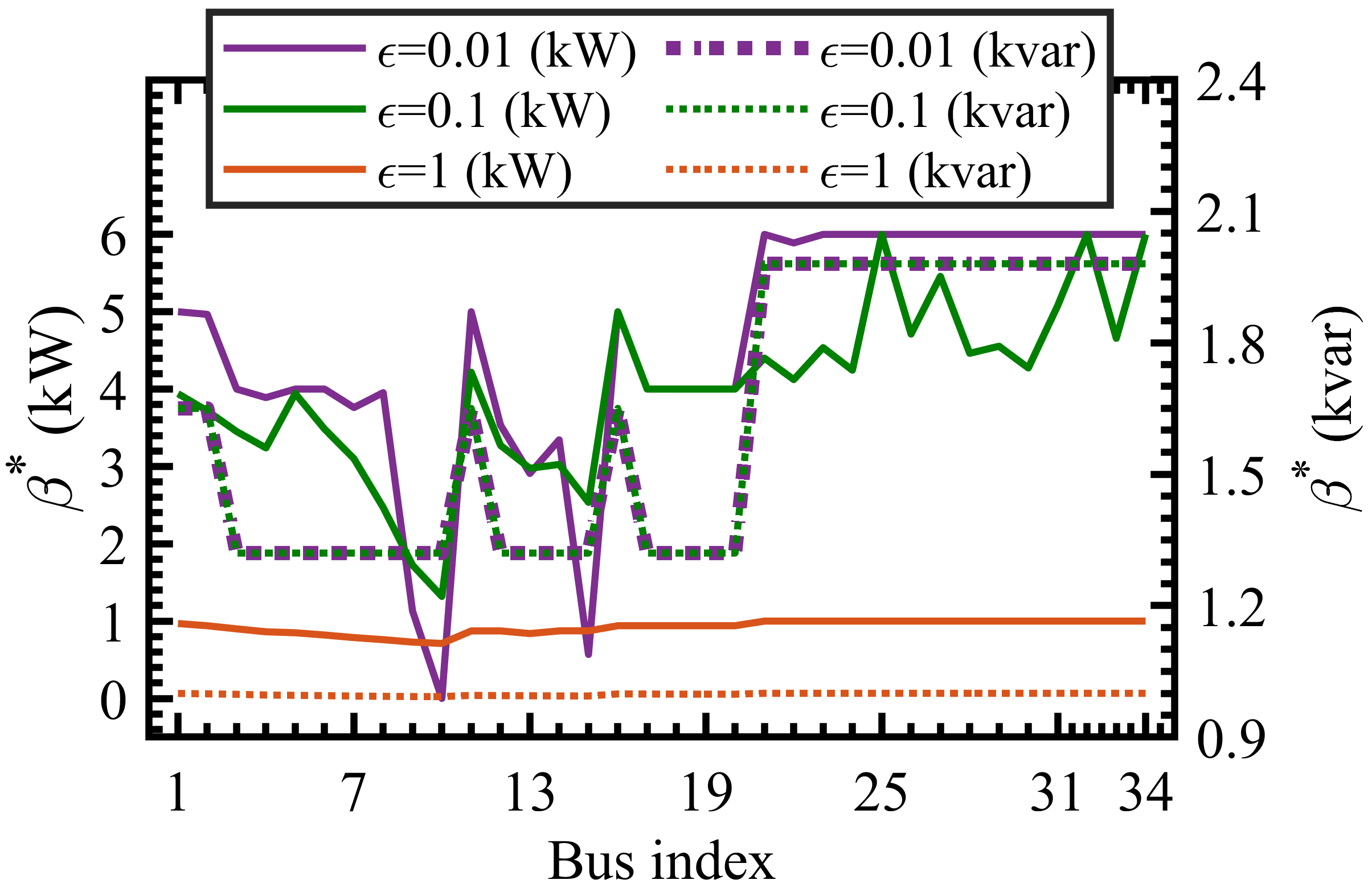}
    \caption{Flexibility assigned to the active and reactive power DOEs for each prosumer at $12{:}00$.}
    \label{bet_145}
\end{figure}

The DOE limits determined by the proposed approach for different values of the design parameter $\epsilon$ are illustrated. Note that $\epsilon$ regulates the trade-off between optimality and flexibility, as discussed in Section~\ref{sec: FOM}. The DOE is computed for three different values of $\epsilon$: $0.01, 0.1$ and $1$, while the scalar weight $w_i$ is uniformly set to $10$ based on manual calibration to ensure balanced flexibility across all prosumers $i\in \mathcal{P}$. The resulting DOE limits are analyzed for the interval beginning at $12{:}00$, a period characterized by significant PV generation, and are illustrated in Fig. \ref{eps}. For a given preferred active power exchange $P^{\mathtt{inj},\star}$, the upper and lower DOE limits for active and reactive power are presented in the left and middle panels of Fig. \ref{eps}, respectively. Additionally, the optimal flexibility parameters $\beta^\star$, which quantify the allowable deviations around the nominal injection values, are shown in Fig. \ref{bet_145}. The corresponding maximum and minimum bus voltage magnitudes, derived from the upper and lower DOE limits, respectively, are depicted in the right panel of Fig. \ref{eps}.

For buses $1$–$20$, reducing $\epsilon$ (from $1$ to $0.1$ and $0.01$) increases the flexibility margin, resulting in wider DOE limits at buses closer to the substation. However, this increased flexibility propagates voltage rise and drop issues toward the end-feeder buses, particularly buses $10$ and $15$, as shown in right panel of Fig. \ref{eps}. Consequently, these buses receive limited flexibility and narrower DOE limits, potentially leading to PV curtailment when BESS and load flexibility are insufficient. In contrast, for buses $21$–$34$, decreasing $\epsilon$ consistently enlarges the flexibility margins without inducing voltage violations. For these buses, the upper DOE limits approach the maximum allowable injection of $6$ kW for $\epsilon=0.1$ and $0.01$, while the lower DOE limits also approach this bound for $\epsilon=0.01$. Moreover, the flexibility margin $\beta^\star$ (Fig. \ref{bet_145}) is larger for $\epsilon=0.01$ and $0.1$ than for $\epsilon=1$. Correspondingly, the reactive power DOE limits (middle panel of Fig. \ref{eps}) for $\epsilon=0.01$ and $0.1$ coincide and are wider than those obtained with $\epsilon=1$. 

In contrast, $\epsilon = 1$ yields lower flexibility in prosumer power exchange but avoids voltage issues at the end-feeder buses. Moreover, the upper DOE limits for these buses remain closer to the desired power exchange, substantially reducing the likelihood of PV curtailment. Although voltage magnitudes stay within the permissible range of $0.95$–$1.05$ pu for all cases, the voltages corresponding to $\epsilon = 1$ are closer to the nominal value of $1$ pu. For these reasons, $\epsilon = 1$ is selected for all subsequent analyses in this paper to demonstrate the effectiveness of the proposed flexible DOE framework.

\subsection{Impact of uniform and non-uniform weighting on the proposed DOE}

\begin{figure}
    \centering
    \includegraphics[width=0.75\linewidth]{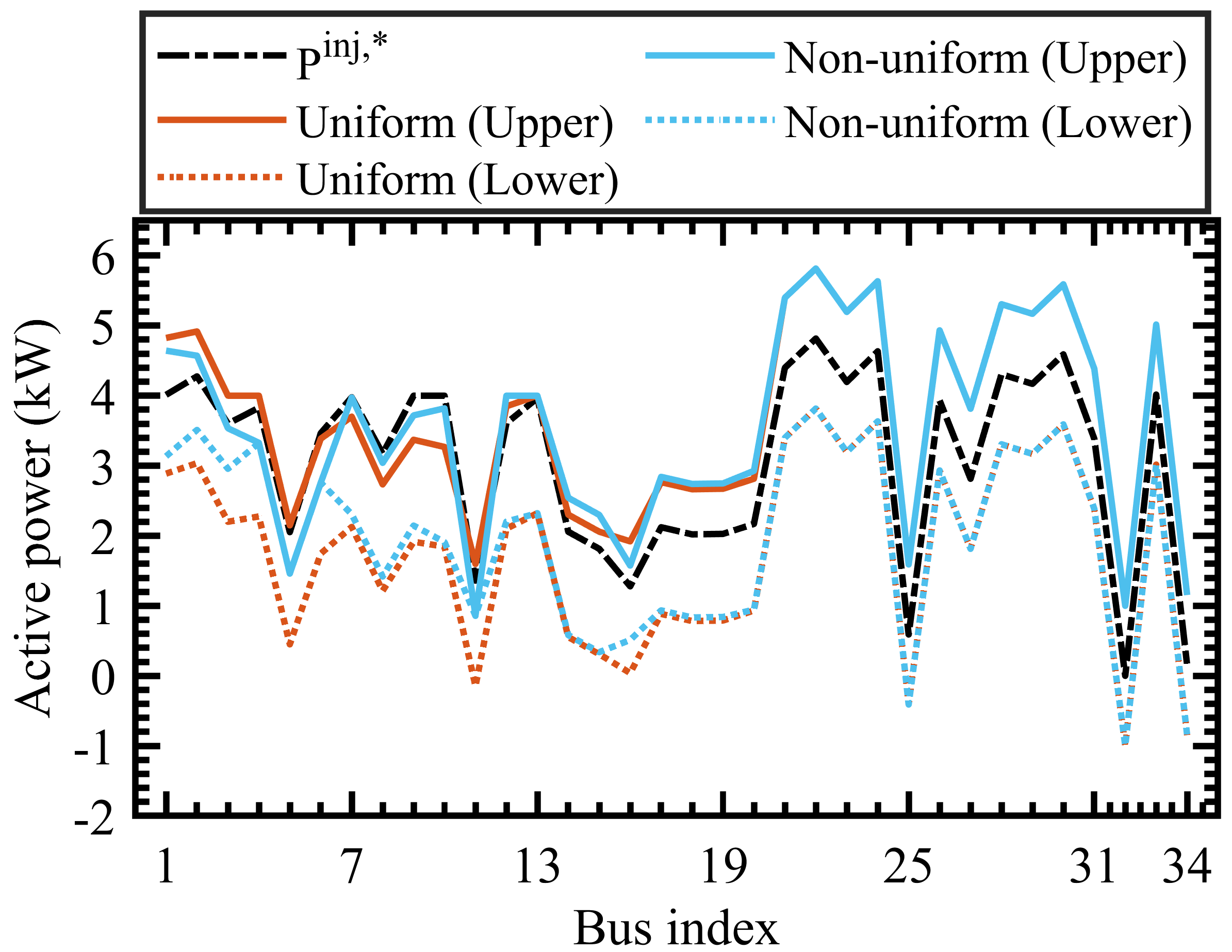} 
    \caption{Upper and lower DOE limits at $12{:}00$ for prosumers' active power, based on preferred setpoint ($P^{\mathtt{inj},\star}$), illustrating the impact of uniform and non-uniform weighting $w_i$ on the proposed DOE.}
    \label{fig:weight}
\end{figure}

\begin{figure*}[h]
    \centering
    \includegraphics[width=0.8\linewidth]{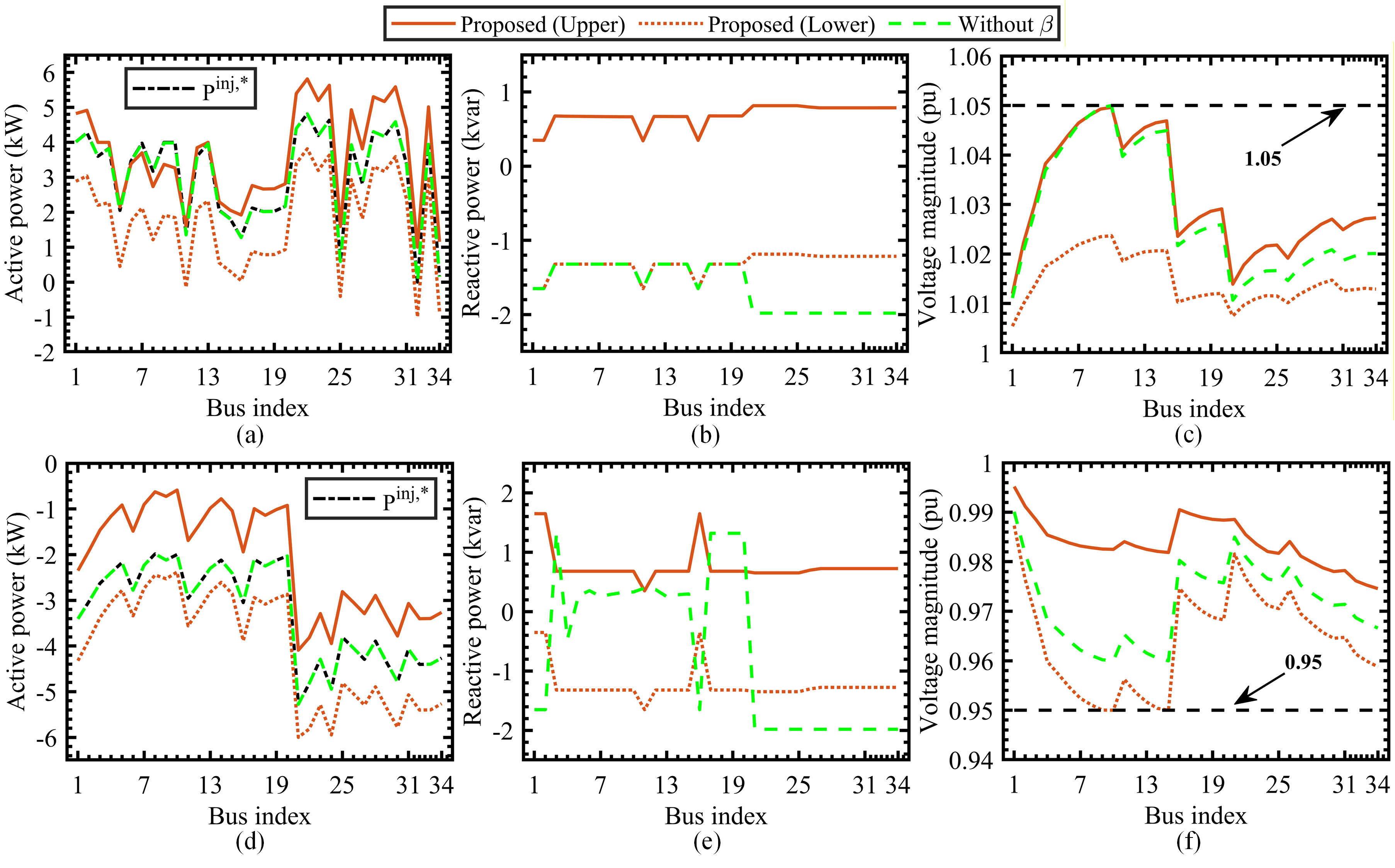} 
    \caption{Comparison of prosumers' active/reactive power DOE limits using the proposed flexible (with $\beta$) and non-flexible (without $\beta$) methods for a given power exchange preference ($P^{\mathtt{inj},\star}$): (a)-(b) at $12{:}00$, (d)-(e) at $20{:}45$; corresponding bus voltage magnitudes: (c) at $12{:}00$ and (f) at $20{:}45$.}
    \label{DOE}
\end{figure*}

A comparison is carried out between flexible DOE limits computed using uniform and non-uniform values of weight $w_i$ for the prosumers $i\in \mathcal{P}$. In the uniform case, all prosumers are assigned $w_i = 10$. In the non-uniform case, distance-aware weights are applied based on the electrical distance from the substation, with higher weights assigned to end-feeder buses and lower weights to buses closer to the substation. Specifically, $w_i = 50$ for end-feeder buses $i \in \{10,14,15,20,25,30,33,34\}$, $w_i = 10$ for buses $i \in \{7,8,9,12,13,17,18,19,24,29,32\}$, and $w_i = 1$ for the remaining buses $i \in \{1,2,3,4,5,6,11,16,21,22,23,26,27,28,31\}$. 

The flexible DOE limits for both weighting strategies are computed for the interval starting at $12{:}00$ and are shown in Fig. \ref{fig:weight}. It is observed from Fig. \ref{fig:weight} that under the non-uniform weighting scheme, buses $1$ through $20$ that are assigned higher weights ($w_i = 10$ or $50$) are granted wider active power DOE limits compared to the uniform weighting scenario. This increased flexibility is achieved at the expense of buses with $w_i = 1$, whose DOE limits are significantly tightened to maintain voltage feasibility. As a result, buses $4$, $5$, $6$, and $11$ receive no active power flexibility, with their upper and lower DOE limits coinciding. In contrast, for buses $21$ through $34$, the DOE limits are nearly identical in both cases due to the absence of voltage issues. These results indicate that while distance-aware weighting enhances flexibility for end-feeder buses, it can entirely eliminate flexibility for prosumers assigned lower weights. Consequently, a uniform weighting scheme with $w_i = 10$ is adopted for all subsequent results.

\subsection{Comparison of the proposed flexible DOE with non-flexible DOE}

The proposed flexible optimization-based DOE is compared with a baseline approach that excludes the flexibility variable $\beta$, resulting in non-flexible DOE limits. This baseline method follows a similar formulation to equation \eqref{eq:dno_optimization}, except for the omission of the $\beta$ variable. To conduct the comparison, two representative time intervals are selected:
\begin{itemize}
    \item $12{:}00$ hour: during peak PV generation, when prosumers are predominantly exporting power to the ADN;
    \item $20{:}45$ hour: during peak evening load, when most prosumer batteries are depleted, leading to net power import from the ADN.
\end{itemize}
The outcomes are compared in Fig. \ref{DOE}. 

At $12{:}00$, Fig. \ref{DOE}a and Fig. \ref{DOE}b present the DOE bounds for active and reactive power exchange, respectively, under both the proposed flexible and the non-flexible methods. The corresponding bus voltage magnitudes for these cases are shown in Fig. \ref{DOE}c. As evident from Fig. \ref{DOE}a, the non-flexible DOE limits (without $\beta$) coincide with the prosumers' preferred injections. In contrast, the proposed method introduces upper and lower flexibility margins around these preferences, offering more leeway for real-time deviations. For prosumers $5$, $6$, and $13$, the upper DOE limits closely match the reference values, whereas for prosumers $7$ through $10$ the upper limits are lower than the reference due to voltage constraints. This behavior is reflected in Fig. \ref{DOE}c, where voltage magnitudes for buses $5$ to $15$ exceed $1.04$ pu and approach the upper regulatory limit of $1.05$ pu at end-feeder buses $9$ and $10$. Any further increase in injection at these locations would lead to voltage violations. Despite these localized restrictions, the overall maximum export capacity across the network at this time is about $123$ kW for the proposed method, representing a $17.52\%$ increase compared to the $104.66$ kW limit in the non-flexible case. This expanded headroom enables prosumers to sell more power, increasing potential revenue while simultaneously reducing the need for curtailment that is subsequently demonstrated.

At $20{:}45$, similar analyses are performed, as shown in Fig. \ref{DOE}d and Fig. \ref{DOE}e, where all prosumers are operating as buyers. The flexible DOE approach once again provides power exchange bounds, whereas the non-flexible method adheres closely to the preferred prosumer injections. As illustrated in Fig. \ref{DOE}f, voltage magnitudes at end-feeder buses $9$, $10$, $14$, and $15$ approach the lower limit of $0.95$ pu. Under these conditions, the proposed method allows for an overall import capacity of $137$ kW, which is about $24.14\%$ higher than the $110.36$ kW limit under the non-flexible approach. In summary, the proposed flexible DOE approach provides wider operational margins for both import and export scenarios, while still respecting network constraints such as voltage limits.

\subsection{Performance of flexible DOE under uncertainty}
\label{subsec: performance}

\begin{figure*}
\centering
   \begin{subfigure}[b]{0.45\textwidth}
            \centering
            \includegraphics[width=0.9\textwidth]{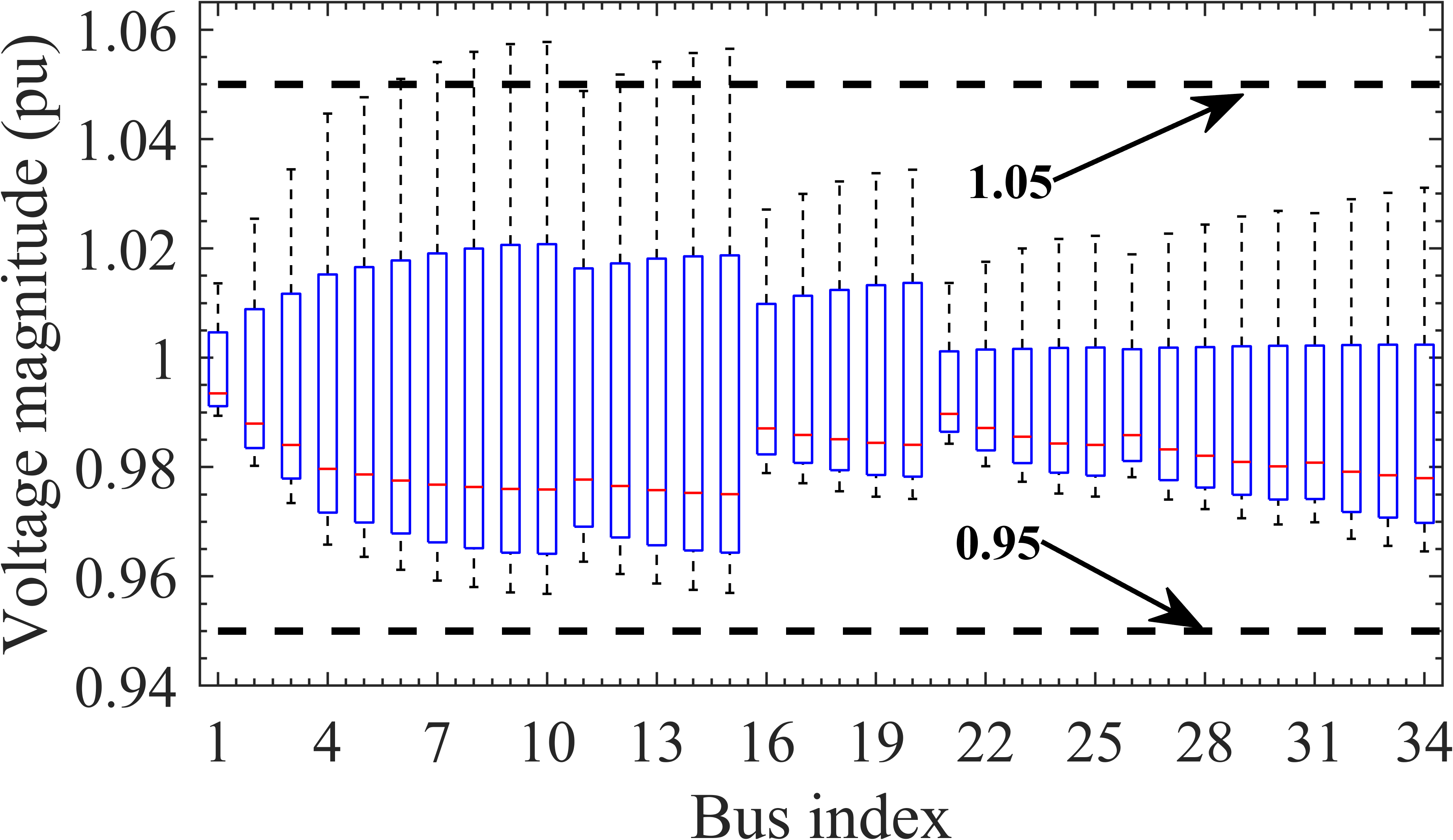}
    \end{subfigure}
    \begin{subfigure}[b]{0.45\textwidth}
            \centering
            \includegraphics[width=0.9\textwidth]{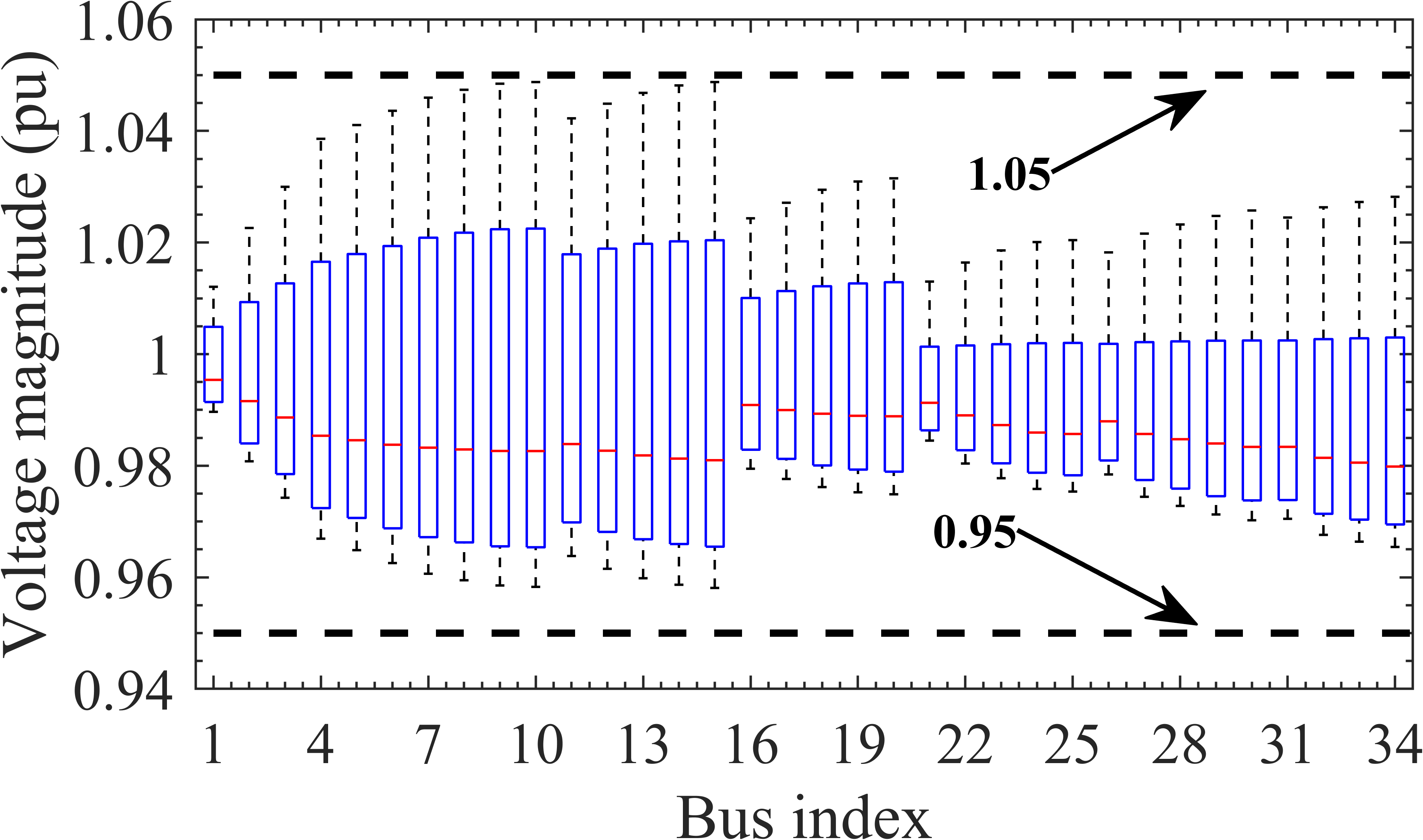}
    \end{subfigure}
    \caption{Box plot of bus voltage profiles: (left) without DOE, (right) with proposed DOE.}\label{fig:DOE_perf}
\end{figure*}
To quantify the benefits of the proposed approach under uncertain PV generation and load demand, the actual PV generation and load are generated by modeling the forecasting errors as uniformly distributed within $\pm 10\%$ of the forecasted values. The performance of the proposed method under these uncertain conditions is evaluated in terms of voltage regulation and PV and load curtailment, as discussed below.

\subsubsection{Voltage magnitude}

Once the actual PV generation and load demand are realized, the prosumer voltage profiles are evaluated under two operating scenarios: (i) power exchange determined without enforcing DOE limits, allowing prosumers to operate up to their injection limits, and (ii) power exchange obtained by re-optimizing under DOE constraints, as formulated in \eqref{eq:pros_opt_post_doe}. The resulting voltage magnitude distributions are presented as box plots in Fig. \ref{fig:DOE_perf}. In each box plot, the box represents the interquartile range (IQR) between the $25^{\text{th}}$ and $75^{\text{th}}$ percentiles, the line inside the box marks the median, and the whiskers extend to capture the full voltage range. In the absence of DOE constraints (left panel of Fig. \ref{fig:DOE_perf}), the voltage magnitude exceed the allowable bounds for the end-feeder buses on several occasions, indicating potential over-voltage violations. In contrast, with the proposed DOE-based scheduling (right panel of Fig. \ref{fig:DOE_perf}), all voltages remain within the safe operating range of $0.95$-$1.05$ pu, demonstrating improved voltage regulation.

\subsubsection{Curtailment of load and PV power}
\label{subsubsec: curt}

\begin{figure}
    \centering
    \includegraphics[width=0.8\linewidth]{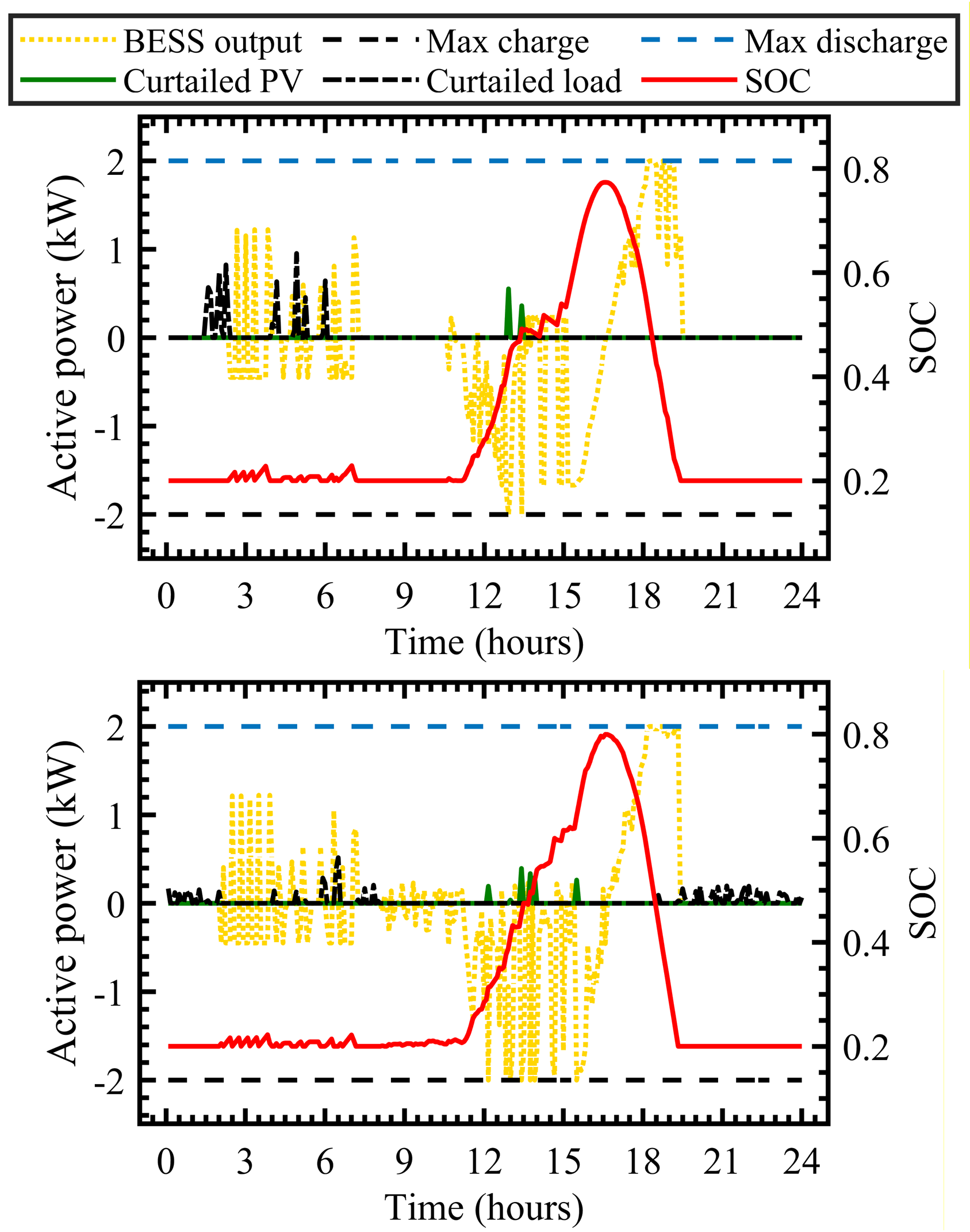} 
    \caption{BESS power output (positive for discharge), SOC, PV/load curtailment, and charge/discharge limits over the day for prosumer $10$ under the (top) proposed flexible DOE method, and (bottom) the non-flexible DOE method.}
    \label{Pbatcurt}
\end{figure}

Real-time deviations from PV generation and load forecasts can force prosumers to curtail PV output or reduce load to stay within the DOE limits, especially when the BESS reaches its operational limits and is unable to absorb surplus PV or supply additional load. This outcome is illustrated for prosumer $10$ in Fig. \ref{Pbatcurt}, which presents the BESS output, SOC, charging/discharging limits, and PV/load curtailments under the proposed flexible DOE method (top panel) and the non-flexible DOE method (bottom panel). As observed, PV curtailment occurs in both cases during periods of high PV generation when the BESS is charging at its maximum rate, leaving excess PV unabsorbed. Similarly, load curtailment in both the cases occurs during morning hours when the BESS SOC is at its minimum limit ($0.2$), preventing the load support. In the non-flexible case, additional load curtailment is observed during evening hours after the BESS is fully discharged while supplying peak demand. Notably, Fig. \ref{Pbatcurt} confirms that all curtailments occur only when the BESS operates at its SOC or power limits.

\begin{figure}
\centering
   \begin{subfigure}[b]{0.45\textwidth}
            \centering
            \includegraphics[width=\textwidth]{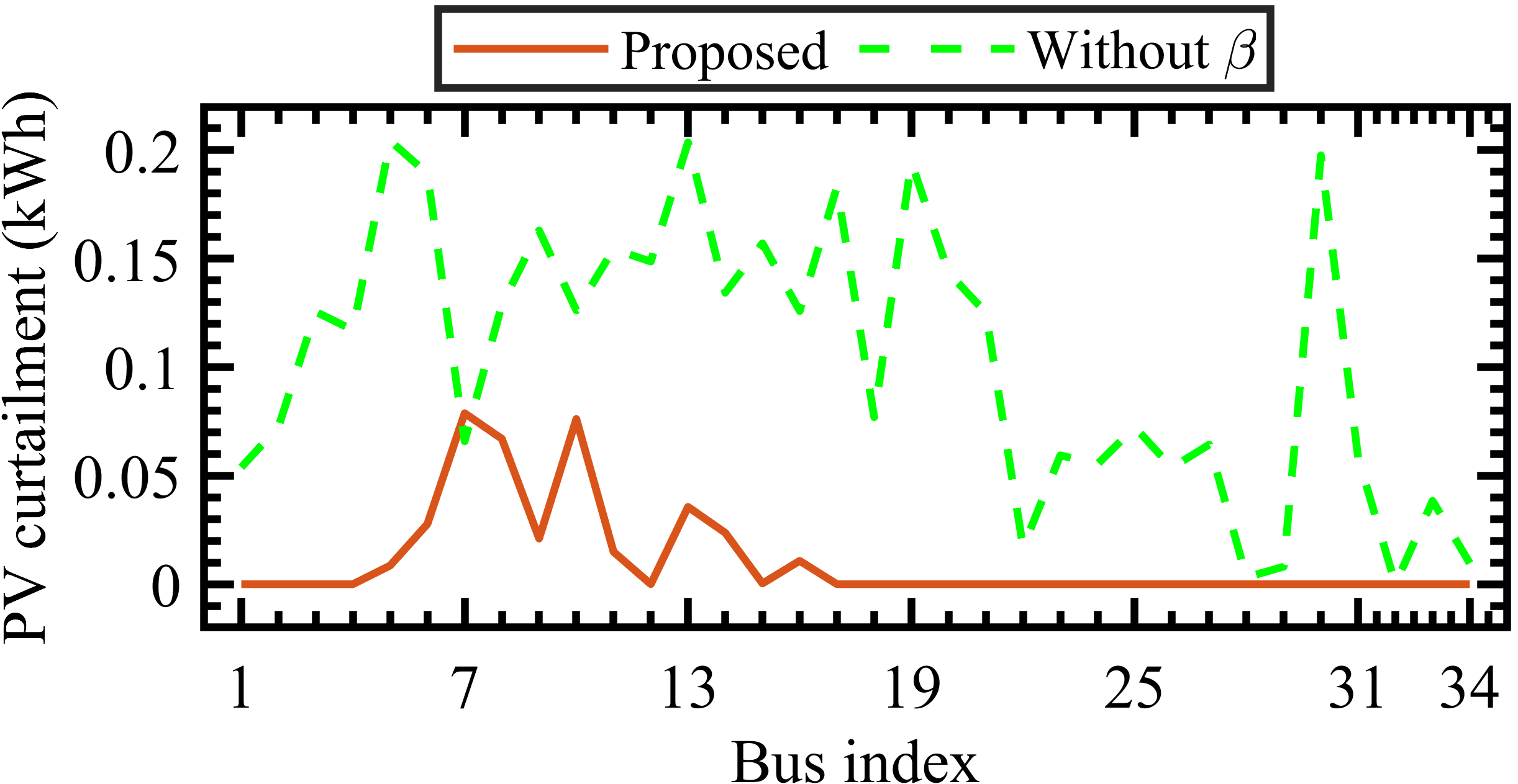}
    \end{subfigure}
    \begin{subfigure}[b]{0.45\textwidth}
            \centering
            \includegraphics[width=\textwidth]{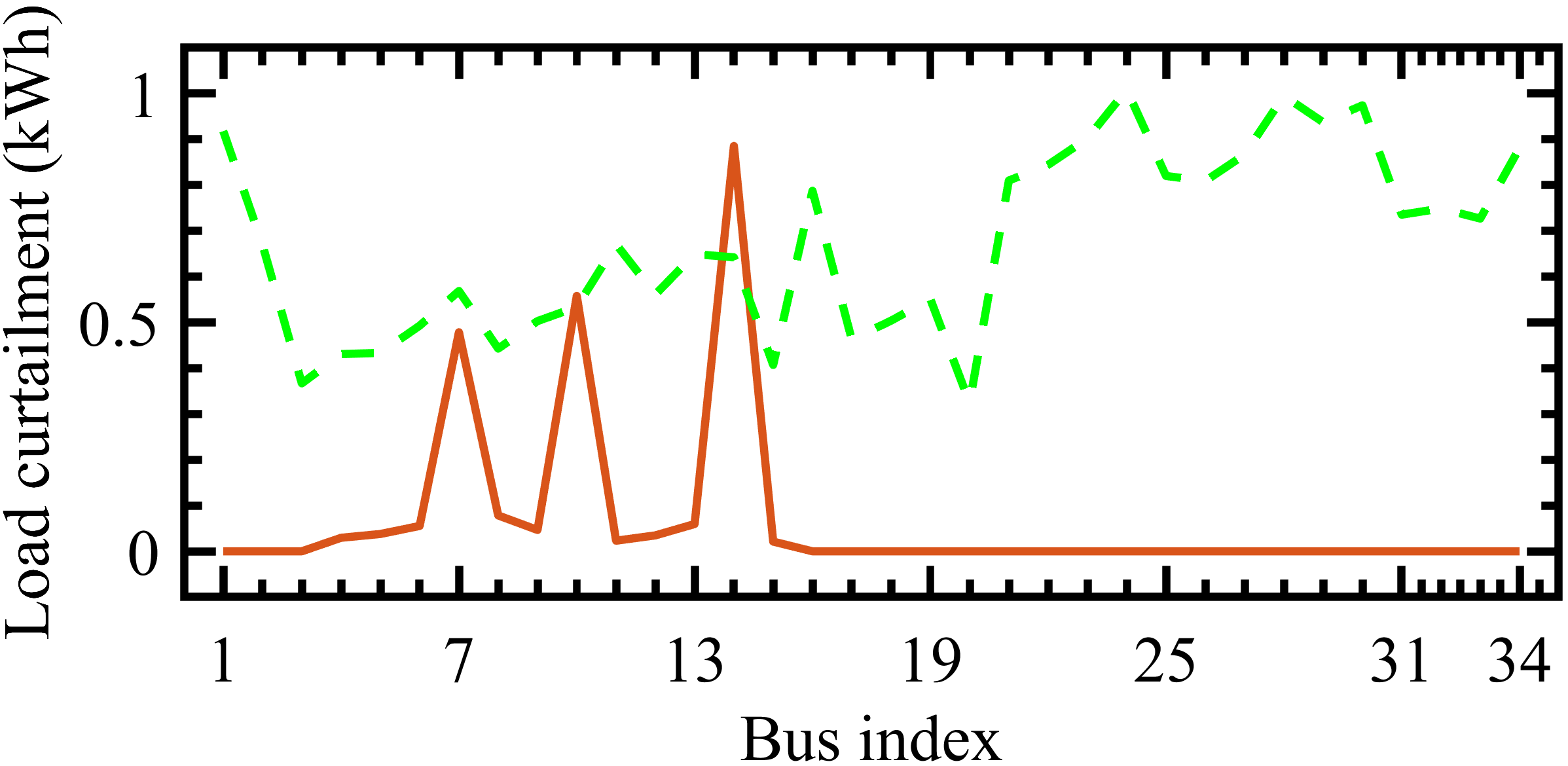}
    \end{subfigure}
    \caption{Aggregate curtailment across prosumers over a day using proposed flexible and non-flexible (without $\beta$) methods: (top) PV curtailment, (bottom) active load curtailment.}\label{fig:curt}
\end{figure}

Figure \ref{fig:curt} compares the total PV and load curtailment under the proposed flexible and non-flexible DOE approaches. PV and load curtailment costs are selected as $c^{\mathtt{PV}}_t=10\pi ^{\mathtt{FiT}}_t$ and $c^{\mathtt{load}}_t=10\pi ^{\mathtt{ToU}}_t$, respectively. The results highlight that the proposed approach significantly reduces curtailment, primarily due to the wider DOE limits enabled by flexibility. Minor exceptions are observed: PV curtailment is slightly higher at bus $7$ because its upper DOE limit is lower than that of the non-flexible approach for most intervals, while load curtailment is marginally higher at bus $10$ and noticeably higher at bus $14$. This occurs because simultaneous battery charging during morning hours narrows the DOE limits for end-feeder buses, and for bus $14$, the lower bound of the flexible DOE is often more restrictive than the DOE limit in the non-flexible approach. Overall, total PV curtailment drops to $0.36$ kWh from $3.53$ kWh, a $89.8\%$ reduction, allowing prosumers to capture greater revenue from excess generation. Similarly, overall load curtailment is reduced from $22.98$ kWh to $2.3$ kWh, a $90\%$ decrease, enhancing prosumer comfort by better accommodating demand variability. 

\subsection{Comparison of operational costs}

\begin{figure}
    \centering
    \includegraphics[width=0.95\linewidth]{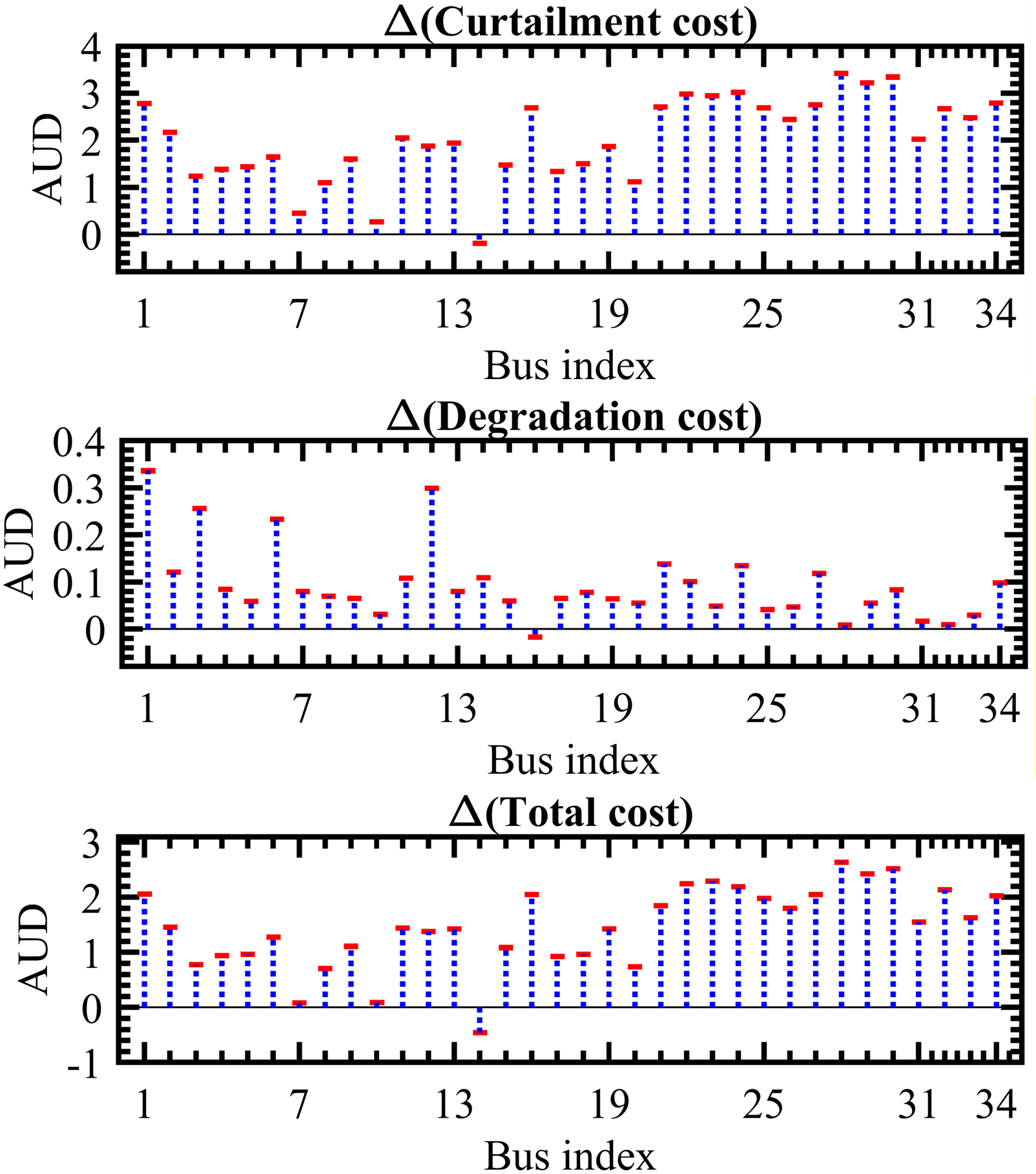} 
    \caption{Difference in aggregate daily cost for each prosumer under the non-flexible and proposed flexible approaches: (top) total PV and load curtailment cost; (middle) BESS degradation cost; and (bottom) total operational cost.}
    \label{fig:cost}
\end{figure}

The aggregate daily operational costs for prosumers are compared between the proposed flexible DOE and the non-flexible DOE approach. The operational cost includes energy purchasing cost, PV and load curtailment costs, battery degradation cost, and revenue from energy sales. The total operational cost is computed as the sum of purchasing, degradation, and curtailment costs minus the revenue. Fig. \ref{fig:cost} illustrates these comparative results by showing the cost difference, defined as $\Delta$(.) and computed as the cost or profit under the non-flexible DOE approach minus that under the proposed flexible DOE approach for each category; therefore, positive values indicate cost savings achieved by the proposed framework.

As shown in Fig. \ref{fig:cost}(top panel), the flexible DOE approach results into significant reduction in the curtailment costs, including both PV and load curtailment, consistent with the findings in Section~\ref{subsubsec: curt}. Furthermore, Fig. \ref{fig:cost}(middle panel) reveals lower battery degradation costs, suggesting reduced stress on the BESS. Finally, Fig. \ref{fig:cost}(bottom panel) demonstrates that, when aggregated across all prosumers, the proposed flexible DOE framework achieves an overall reduction of $11.58\%$ in total operational costs compared to the non-flexible approach. All prosumers except one report a reduction in total operational cost under the proposed approach. In summary, the flexible DOE approach provides clear economic benefits to prosumers by reducing overall operational costs through lower PV and load curtailment, reduced battery degradation, and enhanced flexibility in power exchange.

\subsection{Scalability study}

\begin{figure*}
    \centering
    \includegraphics[width=0.85\linewidth]{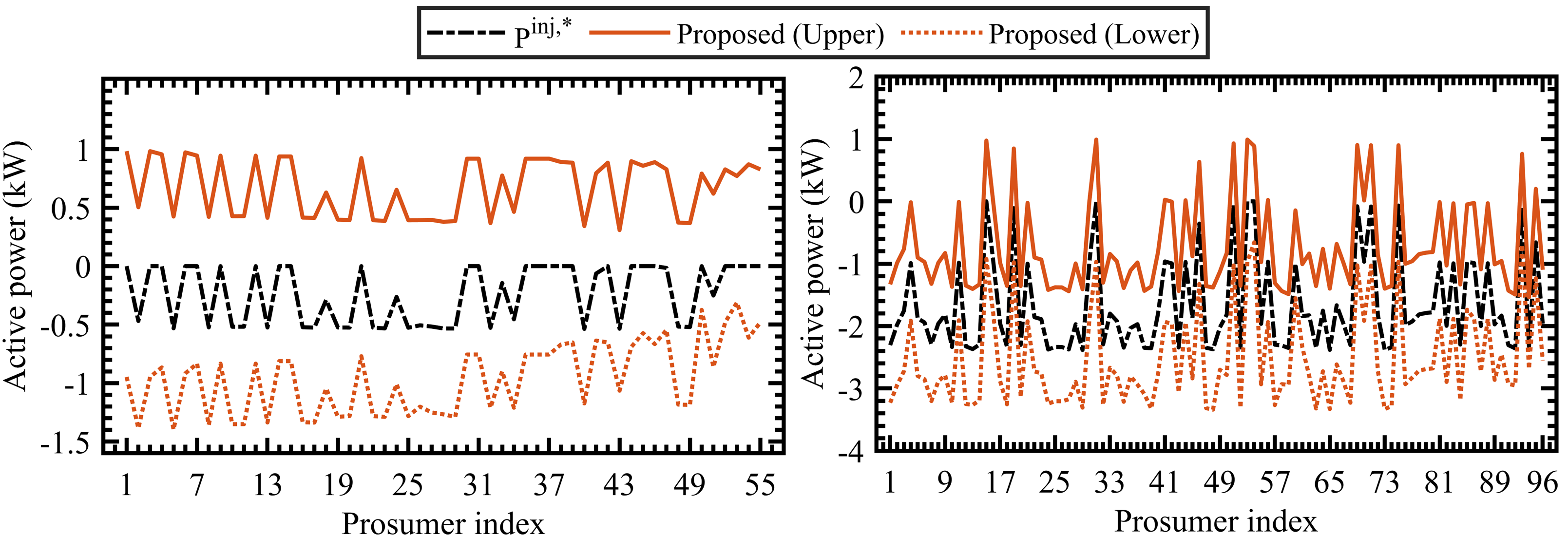} 
    \caption{Upper and lower DOE limits for active power for each prosumer at $20{:}45$ and the preferred setpoint ($P^{\mathtt{inj},\star}$) for the (left) $116$-bus network and (right) $236$-bus network.}
    \label{fig: doe_scal}
\end{figure*}

\begin{figure*}
    \centering
    \includegraphics[width=0.85\linewidth]{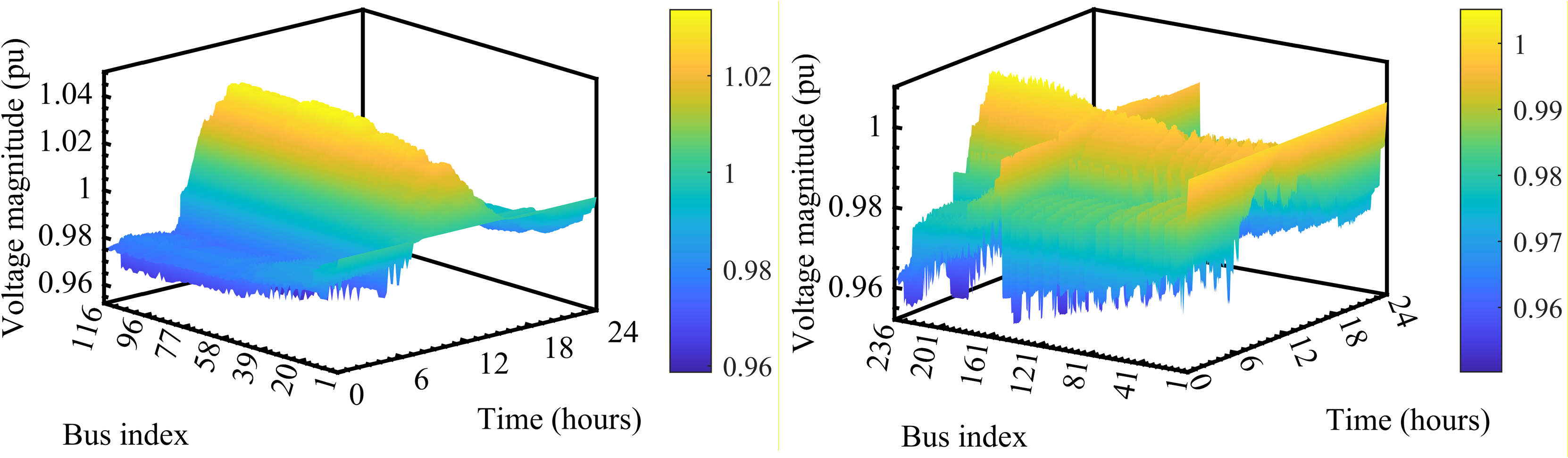} 
    \caption{Voltage magnitude profiles across all buses for the (left) $116$-bus network and (right) $236$-bus network.}
    \label{fig: scal_volt}
\end{figure*}

The scalability of the proposed approach is examined on two larger distribution networks: a $116$-bus system \cite{khan2022reduced} and a $236$ bus system \cite{faia2020optimal}. The $116$-bus network comprises $55$ prosumers, while the $236$-bus network includes $96$ prosumers. The upper and lower DOE limits computed using the proposed flexible method at $20{:}45$,shown in Fig. \ref{fig: doe_scal}, indicate that prosumers are provided with flexible operating bounds centered around their desired power exchange. In addition, voltage profiles for all buses are evaluated under uncertain PV generation and load demand using the flexible DOE limits provided by the DNO. These profiles, presented in Fig. \ref{fig: scal_volt}, confirm that voltage magnitudes across the networks remain within specified bounds. 

The computational effort required by the DNO to determine the DOE is also compared for the $35$-bus, $116$-bus, and $236$-bus networks, as reported in Table \ref{Tab:comp_time}. While computation time per interval increases with network size, it remains limited to $1.12$ seconds for the $236$-bus system. In practice, DNOs typically have access to more powerful computational resources, which would further reduce this time. These results confirm that the proposed approach is scalable and suitable for real-time implementation on large-scale distribution networks.

\begin{table}
\caption{DOE computation time}
  \label{Tab:comp_time}
  \centering
  \footnotesize
  \begin{tabular}{|c|c|c|c|}
    \hline
    Distribution network                & $35$-bus    & $116$-bus          &    $236$-bus                  \\
   \hline
   Time (seconds)                     &    $0.23$     &     $0.52$      &              $1.12$                     \\
    \hline
    \end{tabular}
\end{table}

\section{Summary and conclusion}
\label{Sum_Con}

This work proposes the computation of flexible DOEs using a flexible optimization framework at prosumer nodes within an ADN. The approach incorporates local decision-making, where prosumers determine their preferred active power exchange setpoints based on forecasted PV generation and load data. These setpoints are communicated to the DNO, who in turn calculates prosumer-centric flexible DOEs by considering the network constraints and available capacity and subsequently returns them to the prosumers, enabling them to schedule their DERs while accounting for forecast uncertainty and adhering to network operating limits. 

The framework is validated on a modified Australian low-voltage distribution network, demonstrating its effectiveness in securely integrating high levels of DERs while respecting voltage constraints. Compared to a non-flexible DOE scheme, the proposed approach achieves substantial reductions in curtailment, with load and PV curtailment reduced by $90\%$ and $89.8\%$, respectively. In addition, the framework achieves an overall operational cost reduction of $11.58\%$, mainly due to lower curtailment costs, reduced battery degradation, and increased flexibility in power import and export. The scalability analysis further indicates that the proposed framework can be implemented in real time on larger distribution networks.

Future work will focus on extending the framework to facilitate enhanced prosumer participation in local electricity markets, as well as on developing incentive mechanisms that promote truthful reporting of desired power exchange setpoints.

\section*{Acknowledgments}
This work was supported in part by the Government of India under the Scheme for Promotion of Academics and Research Collaboration (SPARC) (Grant No. SPARC/2019-2020/P1599/SL) at IIT Kanpur and in part by the Science and Engineering Research Board, Government of India, through the sponsored Center of Excellence on Energy Aware Urban Infrastructure, IIT Kharagpur, under Grant IPA/2021/000081. Abhishek Mishra gratefully acknowledges the support by Prime Minister's Research Fellowship.

\appendix

The explicit forms of the coefficients $A^i_t$ and $b^i_t$, $i\in\{1,2,3\}$ from \eqref{power_flow} are provided here. Let $N_\mathtt{line}$ denote the number of lines in the ADN, while, number of buses are represented by $N_\mathtt{bus}$. The substation bus (or node) is indexed as bus $0$ while the remaining buses are index from $1$ to $(N_\mathtt{bus}-1)$. The matrix $\mathbf{\Gamma}$ is of order $((N_\mathtt{bus}-1) \times (N_\mathtt{bus}-1))$, with diagonal entries $(\mathbf{\Gamma})_{ii}=-1$ for all $i \in \{1,2,\ldots,(N_\mathtt{bus}-1)\}$, and with off-diagonal entries $(\mathbf{\Gamma})_{ij}=1$ whenever $j$ is a child node of node $i$, while all remaining entries are zero. Similarly, the matrix $\mathbf{\Lambda}$ is of order $\big(N_\mathtt{bus}\times N_\mathtt{bus}\big)$, where $(\mathbf{\Lambda})_{ii}=-1$ for all $i \in \{1,2,\ldots,N_\mathtt{bus}\}$, and $(\mathbf{\Lambda})_{(i+1),(j+1)}=1$ for every $i \in \{1,2,\ldots,(N_\mathtt{bus}-1)\}$ with $j$ being the parent node of node $i$, and the rest of the entries are zero. Furthermore, the matrix $\mathbf{P_x}$ is of dimension $(N_\mathtt{line} \times 2N_\mathtt{bus})$, with $(\mathbf{P_x})_{i,(i+1)}=1$ for all $i \in \{1,2,\ldots,(N_\mathtt{bus}-1)\}$, and all other entries equal to zero; whereas the matrix $\mathbf{Q_x}$, also of order $(N_\mathtt{line} \times 2N_\mathtt{bus})$, has $(\mathbf{Q_x})_{i,(i+1+N_\mathtt{bus})}=1$ for all $i \in \{1,2,\ldots,(N_\mathtt{bus}-1)\}$ and zeros elsewhere. In addition, the matrix $\mathbf{V}$ is of size $((N_\mathtt{bus}-1) \times N_\mathtt{bus})$, with $(\mathbf{V})_{i,j+1}=1$ for every $i \in \{1,2,\ldots,(N_\mathtt{bus}-1)\}$ with $j$ being the parent node of node $i$ and all other elements equal to zero. The slack bus voltage magnitude is denoted by $V_{\mathtt{slack}}$. $\mathbb{0}^{N_{\mathtt{line}}}$ denotes a zero column vector of length $N_{\mathtt{line}}$. Furthermore, the matrix $G_t$ is defined as $$G_t= \begin{bmatrix}
        \left(\mathbb{0}^{N_{\mathtt{line}}}\right)^\top \\
        \mathbf{I}_{N_{\mathtt{line}}}
        \end{bmatrix},$$
and vectors of line resistances and reactances are denoted by $r$ and $x$, respectively. Finally, matrix $\mathbf{x}^{\mathtt{mat}}$ is of order $({(2N_{\mathtt{bus}}) \times 2n})$ such that $(\mathbf{x}^{\mathtt{mat}})_{(i+1),j}=1$ and $(\mathbf{x}^{\mathtt{mat}})_{(i+1+N_{\mathtt{bus}}),(j+n)}=1$ when prosumer $j \in \mathcal{P}$ is connected to bus $i \in \{1,2 \dots (N_{\mathtt{bus}}-1)\}$, with all remaining entries equal to zero. Now, let us define
\begin{subequations}
    \begin{align}
         P_t^I=&\text{diag}\left(\left(\text{diag}\left(\mathbf{V} \cdot \bar{V}^\mathtt{sq}_t\right)\right)^{-1} \cdot \bar{P}_t \right),
        \\  Q_t^I=&\text{diag}\left(\left(\text{diag}\left(\mathbf{V} \cdot \bar{V}^\mathtt{sq}_t\right)\right)^{-1} \cdot \bar{Q}_t \right),
        \\V_t^I= &\text{diag}\Big( \left(\text{diag}\left(\text{diag}\left(\mathbf{V} \cdot \bar{V}^\mathtt{sq}_t\right)\cdot \left(\mathbf{V} \cdot \bar{V}^\mathtt{sq}_t\right)\right)\right)^{-1}\cdot  \nonumber \\
        &\left(\text{diag}\left(\bar{P}_{t}\right) \cdot \bar{P}_{t} + \text{diag}\left(\bar{Q}_{t}\right) \cdot \bar{Q}_{t}\right) \Big).
    \end{align}
\end{subequations}
Also, we write
\begin{subequations}
    \begin{align}
        & A_t^{11}= \mathbf{\Gamma}^{-1}\cdot \mathbf{P_x},
        \\& A_t^{22}= \mathbf{\Gamma}^{-1}\cdot \mathbf{Q_x},
        \\& A_t^{33}= 2 \mathbf{\Lambda}^{-1} \cdot G_t \cdot \left(\text{diag}\left(r\right)\cdot A_t^{11} + \text{diag}\left(x\right)\cdot A_t^{22} \right),
        \\& b_t^{11}=\mathbb{0}^{N_{\mathtt{line}}},
        \\& b_t^{22}=\mathbb{0}^{N_{\mathtt{line}}},
        \\& b_t^{33}= 2\:\mathbf{\Lambda}^{-1} \cdot G_t \cdot \left(\text{diag}\left(r\right) \cdot b_t^{11} + \text{diag}\left(x\right)\cdot b_t^{22}\right) \nonumber\\
        &- \mathbf{\Lambda}^{-1}\cdot \begin{bmatrix}
            V_{\mathtt{slack}}^2\\\mathbb{0}^{n}
        \end{bmatrix},
        \\& C_t^1=- \mathbf{\Gamma}^{-1}\cdot \text{diag}\left(r\right),
        \\& C_t^2=- \mathbf{\Gamma}^{-1}\cdot \text{diag}\left(x\right),
        \\& C_t^3=\mathbf{\Lambda}^{-1} \cdot G_t \cdot (2\:\text{diag}\left(r\right) \cdot C_t^1 + 2\:\text{diag}\left(x\right) \cdot C_t^2 \nonumber \\ 
        &- \text{diag}\left( \text{diag}\left(r\right) \cdot r+ \text{diag}\left(x\right) \cdot x  \right)  ).
    \end{align}
\end{subequations}
Furthermore,
\begin{subequations}
    \begin{align}
        & D_t= 2P_t^I\cdot A_t^{11}+2Q_t^I\cdot A_t^{22}-V_t^I\cdot \mathbf{V} \cdot  A_t^{33},
        \\& E_t= 2P_t^I\cdot C_t^1+2Q_t^I\cdot C_t^2-V_t^I\cdot \mathbf{V} \cdot  C_t^3,
        \\& F_t= \bar{I}^{\mathtt{sq}}_t+2 P_t^I \cdot b_t^{11}-2 P_t^I \cdot \bar{P}_t +2 Q_t^I \cdot b_t^{22}-2 Q_t^I \cdot \bar{Q}_t \nonumber \\
        &- V_t^I\cdot \mathbf{V} \cdot  b_t^{33}+ V_t^I \cdot \mathbf{V} \cdot \bar{V}^\mathtt{sq}_t.
    \end{align}
\end{subequations}
Therefore, the coefficients $A^i_t$ and $b^i_t$, $i\in\{1,2,3\}$ can be written as-
\begin{subequations}
    \begin{align}
        & A_t^1=\big[A_t^{11}+C_t^1\cdot \left(\left(\mathbf{I}_{N_\mathtt{line}} - E_t \right)^{-1} \cdot D_t\right)\big] \cdot \mathbf{x}^{\mathtt{mat}},
        \\& A_t^2=\big[A_t^{22}+C_t^2\cdot \left(\left(\mathbf{I}_{N_\mathtt{line}} - E_t \right)^{-1} \cdot D_t\right)\big] \cdot \mathbf{x}^{\mathtt{mat}},
        \\& A_t^3=\big[A_t^{33}+C_t^3\cdot \left(\left(\mathbf{I}_{N_\mathtt{line}} - E_t \right)^{-1} \cdot D_t\right)\big] \cdot \mathbf{x}^{\mathtt{mat}},
        \\& b_t^1=b_t^{11}+ C_t^1 \cdot \left(\left(\mathbf{I}_{N_\mathtt{line}} - E_t \right)^{-1} \cdot F_t\right),
        \\& b_t^2=b_t^{22}+ C_t^2 \cdot \left(\left(\mathbf{I}_{N_\mathtt{line}} - E_t \right)^{-1} \cdot F_t\right),
        \\& b_t^3=b_t^{33}+ C_t^3 \cdot \left(\left(\mathbf{I}_{N_\mathtt{line}} - E_t \right)^{-1} \cdot F_t\right).
    \end{align}
\end{subequations}

\bibliographystyle{IEEEtran}
\bibliography{main.bib}

\end{document}